\documentclass{IEEEtran}
\usepackage{cite}
\usepackage{amsmath,amssymb,amsfonts}
\usepackage{algorithmic}
\usepackage{graphicx}
\usepackage{textcomp}
\usepackage{booktabs}
\usepackage{epstopdf}
\usepackage{array}
\usepackage{pgffor}
\usepackage{bm}
\usepackage{subfigure}
\usepackage{color}
\usepackage{url}
\usepackage{acronym}
\usepackage{tabularx}
\usepackage{algorithm}
\usepackage{xpatch}
\acrodef{em}[EM]{Electromagnetic}
\acrodef{admm}[ADMM]{alternating direction method of multiplier}
\acrodef{isp}[ISP]{inverse scattering problem}	
\acrodef{tv}[TV]{total variation}
\acrodef{pnp}[PnP]{plug-and-play }	
\acrodef{dpnp}[DPnP]{Diffuion PnP}	
\acrodef{dip}[DIP]{deep image prior}
\acrodef{gmr}[GMR]{generative model reparameterization}	
\acrodef{map}[MAP]{maximum a posteriori}
\acrodef{mmse}[MMSE]{minimum mean squared error}
\acrodef{dpnp}[DPnP]{diffusion PnP}	
\acrodef{vae}[VAE]{variational autoencoder}	
\acrodef{sde}[SDE]{stochastic differential equation}	
\acrodef{pcs}[PCS]{proximal consistency sampler}	
\acrodef{dds}[DDS]{denoising diffusion sampler}	
\acrodef{mc}[MC]{Markov chain}
\acrodef{mcmc}[MCMC]{Markov chain Monte Carlo}
\acrodef{ldpnp}[L-DPnP]{latent-\ac{dpnp}}
\acrodef{pdpnp}[P-DPnP]{pixel-based \ac{dpnp}}
\acrodef{rmse}[RMSE]{root mean square error}
\acrodef{ssim}[SSIM]{structural similarity index measure}

\usepackage[colorlinks,citecolor=green,urlcolor=blue,bookmarks=false,hypertexnames=true]{hyperref} 
\usepackage{amsmath}

\floatstyle{ruled}
\newfloat{Architecture}{htbp}{loa}
\floatname{Architecture}{Architecture}

\begin{document}

\title{Plug-and-Play Latent Diffusion for Electromagnetic Inverse Scattering with Application to Brain Imaging}
\author{Rui Guo,  \IEEEmembership{Member, IEEE}, Yi Zhang,  \IEEEmembership{Member, IEEE}, Yhonatan Kvich, \IEEEmembership{Graduate Student Member, IEEE}, \\Tianyao Huang, \IEEEmembership{Member, IEEE}, Maokun Li, \IEEEmembership{Fellow, IEEE}, and Yonina C. Eldar,  \IEEEmembership{Fellow, IEEE}
\thanks{This research was supported by the European Research Council (ERC) under the European Union’s Horizon 2020 research and innovation program (grant No. 101000967), by the Israel Science Foundation (grant No. 3805/21, 536/22) within the Israel Precision Medicine Partnership (IPMP) program, and by the Manya Igel Centre for Biomedical Engineering and Signal Processing. (\textit{Corresponding author: Rui Guo})}
\thanks{Rui Guo, Yi Zhang, Yhonatan Kvich, and Yonina C. Eldar are with Faculty of Math and Computer Science, Weizmann Institute of Science, Rehovot 7610001, Israel. }
\thanks{Tianyao Huang is with the School of Computer and Communication
Engineering, University of Science and Technology Beijing, Beijing 100083,
China.}
\thanks{Maokun Li is with the Department of Electronic Engineering, Tsinghua University, Beijing 100084, China. }}

\maketitle

\begin{abstract}
Electromagnetic (EM) imaging is an important tool for non-invasive sensing with low-cost and portable devices. One emerging application is EM stroke imaging, which enables early diagnosis and continuous monitoring of brain strokes. Quantitative imaging is achieved by solving an inverse scattering problem (ISP) that reconstructs permittivity and conductivity maps from measurements. In general, the reconstruction accuracy is limited by its inherent nonlinearity and ill-posedness. Existing methods, including learning-free and learning-based approaches, fail to either incorporate complicated prior distributions or provide theoretical guarantees, posing difficulties in balancing interpretability, distortion error, and reliability. To overcome these limitations, we propose a posterior sampling method based on latent diffusion for quantitative EM brain imaging,  adapted from a generative plug-and-play (PnP) posterior sampling framework. Our approach allows to flexibly integrate prior knowledge into physics-based inversion without requiring paired measurement-label datasets. We first learn the prior distribution of targets from an unlabeled dataset, and then incorporate the learned prior into posterior sampling. In particular, we train a latent diffusion model on permittivity and conductivity maps to capture their prior distribution. Then, given measurements and the forward model describing EM wave physics, we perform posterior sampling by alternating between two samplers that respectively enforce the likelihood and prior distributions. Finally, reliable reconstruction is obtained through minimum mean squared error (MMSE) estimation based on the samples.
Experimental results on brain imaging demonstrate that our approach achieves state-of-the-art performance in reconstruction accuracy and structural similarity while maintaining high measurement fidelity.

\end{abstract}

\begin{IEEEkeywords}
inverse scattering problems, brain imaging, posterior sampling, latent diffusion, plug-and-play.
\end{IEEEkeywords}

\section{Introduction}

\ac{em} imaging is a crucial sensing modality in diverse fields  due to its non-ionizing nature, material penetration capability, and cost-effectiveness \cite{pastorino2010microwave, li2023applications,tan2020efficient, zhang2025solving, salucci2019instantaneous,bevacqua2021millimeter,10919093}.  Its application scenario includes security\cite{tan2020efficient, zhang2025solving}, non-destructive evaluation \cite{massa2006microwave,zoughi2000microwave}, and biomedical diagnostics \cite{salucci2019instantaneous,bevacqua2021millimeter,10919093}. Recently, it has  emerged as a promising tool for early diagnosis and long-term monitoring of strokes  \cite{hamidipour2018electromagnetic,salucci2019instantaneous}. 
Stroke occurs when blood flow to the brain is disrupted by a blockage (ischemic stroke) or a rupture (hemorrhagic stroke). Rapid identification of stroke types and long-term monitoring are essential for improving outcomes \cite{wey2013review}. Traditional imaging modalities—computed tomography (CT), positron emission tomography (PET), and magnetic resonance imaging (MRI)—have limitations: CT and PET involve ionizing radiation, and MRI lacks portability, limiting its use for bedside monitoring.

The physical foundation of \ac{em} stroke imaging is based on the fact that different types of strokes exhibit distinct electrical properties, such as permittivity and conductivity, compared to normal tissue. As EM waves penetrate the head, they are altered by the spatial variations in electrical property. The externally measured \ac{em} fields thus carry information about the internal electrical properties. 

One way to retrieve the permittivity and conductivity is to formulate an \ac{em} \ac{isp} \cite{chen2018computational}, which quantitatively solves for these parameters according to Maxwell's equations, given the temporal and spatial information of excitation sources and  measurements. The resulting \ac{isp} is highly nonlinear due to multiple scattering among heterogeneous tissues and the skull, and ill-posed due to the limited degrees of freedom of the scattered field \cite{bucci1989degrees}. This leads to unstable solutions sensitive to measurement noises. 

{Classical model-based methods formulate the \ac{isp} as a regularized optimization or a Bayesian inference problem. Gauss-Newton \cite{abubakar2012application}, conjugate gradient \cite{1017641}, and \ac{admm} \cite{liu2020alternating}, are then used to iteratively update the unknowns. These approaches face limitations of non-convex landscapes. When initial guesses deviate significantly from the true solutions, local minima can trap the optimization process. In Bayesian inversion, sampling methods such as \ac{mcmc} are used to sample from assumed distributions, but they are usually more computationally expensive than optimization approaches. Furthermore, both methods use oversimplified priors such as handcrafted regularizers ($\ell_2$ norm or \ac{tv}) or simple distributions \cite{abubakar2012application, 1017641,liu2020alternating, shen2021bayesian, zhang20233,ZhouYin}, which cannot capture realistic structures in complex materials.} To better constrain target shapes, handcrafted reparameterization basis, such as geometry parameters\cite{10919093}, radial basis \cite{li2012three} and wavelets \cite{guo2015microwave}, have been proposed. However, these techniques are tailored to specific scenarios and lack flexibility for broader applications.  

Recently, deep learning methods were suggested to address \ac{isp}s  \cite{chen2020review,li2018deepnis,li2021machine, ye2023deep,li2023applications,guo2023physics}. A straightforward approach is to learn the mapping between measurements and property maps \cite{li2018deepnis}. However, purely data-driven models suffer from several limitations. Their black-box nature lacks physical interpretability, making them less suitable for security/clinical/industrial applications where reliability is crucial. 
Additionally, these methods exhibit setup rigidity—they require fixed measurement configurations during both training and inference, including transmitter/receiver geometries, excitation waveforms, sensor frequency responses, and radiation patterns \cite{geffrin2005free}. This imposes strict consistency requirements on the measurement system, which are often impractical due to varying constraints in real world scenarios. Consequently, any modification to the measurement setup necessitates time-consuming {dataset re-generation and neural network re-training, which involves thousands to millions of computationally intensive full-wave simulations}.

To enhance interpretability and generalizability of data-driven approaches, many works integrate physical models with the deep learning process \cite{li2021machine,ye2023deep,li2023applications}.
These hybrid methods are known to lead to more accurate networks while requiring fewer learned parameters compared to purely data-driven techniques \cite{monga2021algorithm, Shlezinger}.  One branch of this integration is \textit{physics-embedded deep learning} \cite{guo2023physics}, where algorithms incorporate physical principles in different parts of the data-driven workflow, such as designing inputs and labels \cite{wei2018deep,ye2020inhomogeneous,cohen2025deep}, loss functions \cite{jin2020physics,bar2021strong}, and architectures of neural networks \cite{tom,guo2021physics}.  Nevertheless, these methods still inherit the setup rigidity limitations from typical data-driven methods. 
Another promising branch is the \textit{\ac{pnp}} approach based on learned priors, which integrates neural network-expressed prior knowledge into classical physics-driven frameworks \cite{venkatakrishnan2013plug, bora2017compressed}. Although this method has slower online execution, it offers high flexibility by learning priors from image-domain data (e.g., property maps) without being tied to specific measurement setups. The computational cost for training is substantially reduced by eliminating repeated full-wave simulations for each new configuration -- the most time-consuming part in dataset preparation and \textit{physics-embedding deep learning}.  Moreover, since the imaging process remains primarily physics-driven, it generally guarantees high data fidelity, which is a critical criterion for \ac{em} imaging \cite{li2021machine}.

Representative \ac{pnp} methods for \ac{em} imaging, which use neural network parameters to represent electrical property maps, include \ac{dip} \cite{ulyanov2018deep, deepeit} and \ac{gmr} \cite{bora2017compressed,10332205,Khorashadizadeh}. In \ac{dip}, the output of neural networks are considered property maps and connected to the forward modeling function. The neural network parameters are optimized through minimizing the data residual between measurements and forward simulations. Neural network architectures act as implicit regularizers during optimization. In \ac{gmr}, pre-trained generative neural networks that embed complex prior knowledge are used to reparameterize the property maps. Instead of optimizing the whole network in \ac{dip}, the unknowns in \ac{gmr}  are only latent variables of the pre-trained generative models. For example,  in \cite{deepeit}, the authors apply \ac{dip} using an untrained U-net for electrical impedance tomography to produce conductivity maps with sharp boundaries. In \cite{10332205}, \ac{gmr} is used to encode electrical properties into the latent space of a \ac{vae}, which yields clear structural features and improves resolution in microwave brain imaging. Both \ac{dip} and \ac{gmr} treat the inverse problems as deterministic optimization problems, which may get trapped in local minima. 

Another limitation of \ac{dip} and \ac{gmr} is that they are \ac{map}-based methods that reconstruct a single property map maximizing a posterior probability. This may produce overconfident estimates when the posterior distribution is multimodal. Using a \ac{mmse} metric can  avoid overconfident solutions by computing the posterior expectation, which is usually approximated by the sample average of the posterior distribution. 
Advances in diffusion models enable posterior sampling with complex priors and tractable computational costs \cite{kadkhodaie2021stochastic,chung2022diffusion,NEURIPS2022_95504595,song2022solving}. However, most existing works are developed for linear problems, and many of them lack theoretical guarantees. Recently, relying on the generative \ac{pnp} and denoising priors, Bouman and Buzzard proposed an asymptotically exact method for posterior sampling based on iterative sampling \cite{Bouman}. This approach is extend to \ac{dpnp} posterior sampling for general inverse problems with a convergence guarantee \cite{xu2024provably, wu2024principled}. \ac{dpnp} alternatively performs a likelihood step  that generates samples consistent with the measurements, and a denoising diffusion sampler that enforces the prior constraint in the image pixel domain. This method was tested on vision problems, such as phase retrieval, quantized sensing, and super resolution \cite{xu2024provably}, but was not considered to solve \ac{isp}s. Due to the drastically varying sensitivities of the measurement with respect to pixels of the electrical property map, directly applying \ac{pdpnp}  to \ac{isp} has intolerably slow convergence, which limits the practical applicability of this method to the physics-based inversion domain where the computational cost of forward modeling is very high. 

To address the aforementioned challenges, we propose \ac{ldpnp}, {where we aim to compute the \ac{mmse} estimate of the \ac{isp} under a Bayesian setting and achieve posterior sampling in a customized latent space that encodes domain-specific electrical parameter correlations. }
To the best of our knowledge, this is the first \ac{isp} posterior sampling \ac{pnp} framework with learned priors. Our contributions are summarized as follows:
\begin{itemize}
    \item {The proposed \ac{ldpnp} framework enables posterior sampling for a wide range of computational imaging problems, by addressing the slow convergence limitation of \ac{pdpnp} caused by significantly varying sensitivities of unknowns}.
  \item We apply the \ac{ldpnp} framework to quantitative \ac{em} inversion, which improves reconstruction accuracy by incorporating learned prior knowledge and by avoiding overconfident solutions in \ac{map} estimation, and naturally provides uncertainty estimates.  
   \item Through  brain strokes imaging, we show that the proposed approach achieves superior reconstruction accuracy while maintaining high data fidelity comparable to other state-of-the-art baselines.
\end{itemize}
Preliminary results were presented in a conference paper \cite{guoposterior}. Herein, we introduce the derivation of \ac{ldpnp},  evaluate the method under broad testing scenarios, compare it against a wide range of existing techniques, provide a detailed explanation of its advantages, and describe implementation details not covered previously. Results in this paper show high robustness to noise and reconstruction accuracy, indicating its reliability for brain imaging. 

The paper is organized as follows. Section \ref{SecII} introduces preliminary knowledge about the \ac{em} wave physics and formulate the \ac{isp}.  Section \ref{SecIII} establishes the \ac{ldpnp} framework and Section \ref{sec_implementation} introduces detailed posterior sampling algorithms.  In Section \ref{SecV}, we validate our approach under two toy scenarios and apply it to brain imaging, comparing the results with existing methods. Section \ref{sec:discussion} discusses the advantages and limitations of 
\ac{ldpnp}, as well as future directions. Section \ref{SecVI} concludes the work.

\section{Problem formulation}\label{SecII}
In this section, we begin by presenting the measurement setup in our scenario and  introducing the wave physics involved in \ac{em} imaging.  We then formulate our inverse scattering problem that needs to be solved for imaging.

\subsection{Measurement setup and wave physics}\label{forward_model}
\begin{figure}[!]
	\centering
	\includegraphics[width=74mm]{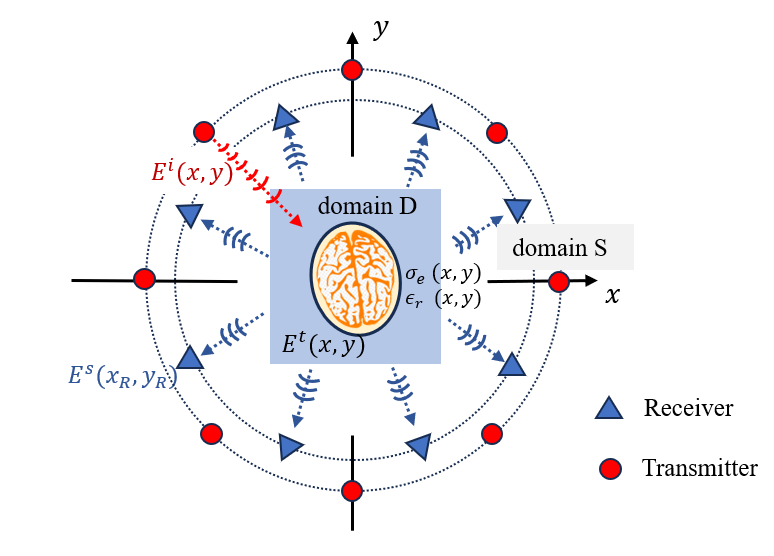}	
	\caption{Illustration of the \ac{em} imaging scenario in a 2-D setup. The transmitters sequentially transmit incident fields $ {E^i}$ into the investigation domain $D$ that is characterized by permittivity and conductivity. The transmitters and receivers are located in domain $S$, where there is no permittivity or conductivity inhomogeneities. }
	\label{fig2}
\end{figure}

We consider a 2-D measurement scenario shown in Fig.~\ref{fig2}, which corresponds to a non-contact sensing scenario where the antennas are  situated in air or an impedance-matched medium. The target of interest, for example, human head,  is located inside $D$, surrounded by \(N_T\) transmitters and \(N_R\) receivers placed in domain \(S\). Domain \(S\) is homogeneous and shares the same electrical properties as the background of \(D\).  

\ac{em} wave propagation is governed by the Maxwell’s equations. The transmitters  illuminate the entire domain with \textit{incident field} $ {E^i}(x,y)$, where $(x,y)$ denotes the coordinate of the 2-D space.  When the target is illuminated, the spatially varying electrical properties—relative permittivity \(\epsilon_r(x,y)\) and conductivity \(\sigma_e(x,y)\)—alters the field distribution,  resulting in a \textit{total field} \(E^t(x,y)\). The receivers collect the \textit{scattered fields} $ {E^s}(x_R,y_R)$  at location $(x_R, y_R)$, which are uniquely determined by the ${E^t}(x,y)$, $\epsilon_r(x,y)$ and $\sigma_e(x,y)$. The scattered fields collected under all transmissions constitute the observation data used for imaging.

We next describe this scattering process  in matrix forms for a single transmission at one frequency. We first discretize domain $D$ into $N_x\times N_y$ subunits, resulting in a discrete representation of relative permittivity and conductivity, i.e.,  $\boldsymbol{\epsilon_r} \in \mathbb R^{N_x\times N_y} $ and  $\boldsymbol{\sigma_e} \in \mathbb R^{N_x\times N_y}$, respectively. Denote the index of transmitter, receiver, and frequency points to be $p$, $q$, and $g$, respectively, with $p=1,...,N_T$, $q=1,...,N_R$, and $g=1,...,N_f$. For the transmitter $p$ at transmitting frequency $f_g$, the \textit{incident field}, \textit{total field} and \textit{scattered field} are denoted by $\mathbf {E^i}_{p,g}\in\mathbb C^{N_xN_y}$, $\mathbf {E^t}_{p,g}\in\mathbb C^{N_xN_y}$ and  $\mathbf {E^s}_{p,g}\in\mathbb C^{N_R}$, respectively. Their relations are given by \cite{jin2015theory,chen2018computational}
\begin{flalign}
& \mathbf{E^t}_{p,g}= \mathbf{E^i}_{p,g} + \mathbf G_D \boldsymbol{\chi} (\boldsymbol{\epsilon_r}, \boldsymbol{\sigma_e}) \mathbf{E^t}_{p,g}, \label{efie}\\
&  \mathbf{E^s}_{{p,g}} = \mathbf G_S \boldsymbol{\chi} (\boldsymbol{\epsilon_r}, \boldsymbol{\sigma_e}) \mathbf{E^t}_{p,g}, \label{data_equation}
\end{flalign}
where $\mathbf{G}_D\in\mathbb C^{N_xN_y\times N_xN_y}$ is the Green's function related to the specific background material and frequency but not related to receiver locations,  and $\mathbf G_S \in \mathbb C^{N_R\times N_xN_y}$ is the Green's function matrix that propagates $\mathbf{E^t}_{p,g}$ to the receivers, and is related to the background material, frequency, and receiver locations. Finally, $\boldsymbol{\chi} (\boldsymbol{\epsilon_r}, \boldsymbol{\sigma_e}) \in\mathbb{C}^{N_x N_y \times N_xN_y}$ is called the contrast matrix, which is a diagonal matrix constructed from $\boldsymbol{\epsilon_r}$ and  $\boldsymbol{\sigma_e}$, with the diagonal given by
\begin{equation}
	\text{diag} \big(\boldsymbol{\chi} (\boldsymbol{\epsilon_r}, \boldsymbol{\sigma_e}) \big) = \text{vec}\bigg\{ \frac{\boldsymbol{\epsilon_r}}{{\epsilon_{r,b}}} - \text{j}\frac{\boldsymbol{\sigma_e}}{2\pi f_g \epsilon_0 \epsilon_{r,b}} -1\bigg\}\in \mathbb C^{N_x N_y}.
\end{equation}
Here ${\epsilon}_{r,b}\in \mathbb C$ is the complex relative permittivity of a known background, $\epsilon_0$ is the permittivity in vacuum, $\text{vec}\{\cdot\}$ is the vectorization operation, and $\text{j}=\sqrt{-1}$ denotes the imaginary unit.
Details about constructing $\mathbf G_D$, $\mathbf G_S$ and $\mathbf{E^i}_{p,g}$ and computing (\ref{efie}) and (\ref{data_equation}) can be found in \cite{chen2018computational}.

For compact representation, we eliminate $\mathbf {E^t}_{p,g}$ by expressing it from (\ref{efie}) and substituting  it into (\ref{data_equation}),  yielding
 \begin{equation}\label{eq:forward}
\mathbf{E^s}_{{p,g}} =  \mathbf G_S \boldsymbol{\chi} (\boldsymbol{\epsilon_r}, \boldsymbol{\sigma_e})\big (\mathbf I - \mathbf G_D \boldsymbol{\chi} (\boldsymbol{\epsilon_r}, \boldsymbol{\sigma_e}) \big)^{-1} \mathbf{E^i}_{p,g},
 \end{equation}
from which we can see that the relationship between $\mathbf{E^s}_{{p,g}}$ and $\boldsymbol{\chi} (\boldsymbol{\epsilon_r}, \boldsymbol{\sigma_e})$ is nonlinear. We then define a nonlinear operator $F_{p,g}(\cdot)$ that maps $\boldsymbol{\epsilon_r}$ and $\boldsymbol{\sigma_e}$ to $\mathbf{E^s}_{{p,g}}$, that is, 
\begin{equation}\label{key}
		F_{p,g} (\boldsymbol{\epsilon_r}, \boldsymbol{\sigma_e})=  \mathbf G_S \boldsymbol{\chi} (\boldsymbol{\epsilon_r}, \boldsymbol{\sigma_e})\big (\mathbf I - \mathbf G_D \boldsymbol{\chi} (\boldsymbol{\epsilon_r}, \boldsymbol{\sigma_e}) \big)^{-1} \mathbf{E^i}_{p,g}.
\end{equation}

During measurement, each transmitter sequentially illuminates domain $D$, and for every transmission, the resulting scattered fields are measured at multiple frequencies. This process yields an observation set comprising all transmitter-receiver-frequency pairs.
By aggregating the scattered field data  together, we have the observation vector  $	\mathbf{d}_{obs}  = \{ \{ \mathbf{E}^{s}_{p,g} \}_{p=1}^{N_T} \}_{g=1}^{N_f} \in \mathbb{C}^{N_T N_R N_f}$. Similarly, we define a nonlinear operator $F(\cdot)$ that simultaneously models the scattered field generated by $\boldsymbol{\epsilon_r}$ and $\boldsymbol{\sigma_e}$ for all transmitters, receivers, and frequencies, that is, $F(\cdot) = \{ \{F_{p,g}(\cdot) \}_{p=1}^{N_T} \}_{g=1}^{N_f}$, where $F(\cdot)$ is first-order differentiable. 
Consequently, the measurement can be modeled as
\begin{equation}\label{original_forward_model}
	\mathbf{d}_{obs}  = F( \boldsymbol{\epsilon_r}, \boldsymbol{\sigma_e}) + \mathbf{n}_{{obs}},
\end{equation}
where $\mathbf{n}_{{obs}}$ denotes measurement noise.
\subsection{Inverse scattering problem}
We wish to solve for $\boldsymbol{\epsilon_r}$ and $\boldsymbol{\sigma_e}$  in (\ref{original_forward_model}) given $\mathbf{d}_{obs}$, from which the physical conditions of  human tissues can be inferred. The inverse problem is severely ill-posed. This lead to numerous pairs of $\boldsymbol{\epsilon_r}$ and $ \boldsymbol{\sigma_e}$ that approximately reproduce  $\mathbf d_{obs}$ \cite{{bucci1989degrees}}, whereas only a limited number of them correspond to the material in our imaging scenario. Therefore, incorporating prior knowledge is essential to eliminate solutions that do not align with real-world feasibility.

We adopt a probabilistic setting that describes the original deterministic \ac{isp} as a Bayesian inverse problem. In this setting, $\mathbf{d}_{{obs}}$, $\boldsymbol{\epsilon_r}$, and $\boldsymbol{\sigma_e}$ are treated as random variables, and prior distributions on $\boldsymbol{\epsilon_r}$ and $\boldsymbol{\sigma_e}$ serve to constrain the solution space. 

Under this Bayesian setting, our goal is to estimate the optimal $\boldsymbol{\epsilon_r}$ and $\boldsymbol{\sigma_e}$ from the observed data $\mathbf{d}_{{obs}}$. The \ac{mmse} estimator is known to be optimal with respect to the $\ell_2$ loss, which is defined as:
\begin{equation} \label{eq:mmse_joint}
	\begin{aligned}
		(\hat{\boldsymbol{\epsilon}}_r, \hat{\boldsymbol{\sigma}}_e)_{\text{MMSE}} &= \arg \min _{(\hat{\boldsymbol{\epsilon}}_r, \hat{\boldsymbol{\sigma}}_e)} \mathbb{E} \left[ \| (\boldsymbol{\epsilon_r}, \boldsymbol{\sigma_e}) - (\hat{\boldsymbol{\epsilon}}_r, \hat{\boldsymbol{\sigma}}_e) \|^2 \right] \\
		& =  \mathbb{E}\left[ (\boldsymbol{\epsilon_r}, \boldsymbol{\sigma_e}) |\mathbf d_{obs}\right].
	\end{aligned}
\end{equation}
The expectation in (\ref{eq:mmse_joint}) is not analytically tractable. We therefore proceed to approximate  it by taking the average of all samples from the posterior  $p(\boldsymbol{\epsilon_r}, \boldsymbol{\sigma_e} | \mathbf{d}_{obs})$:
\begin{equation}\label{mmse_average}
	(\hat{\boldsymbol{\epsilon}}_{\boldsymbol r}, \hat{\boldsymbol{\sigma}}_{\boldsymbol e})_{\text{MMSE}} \approx \frac{1}{M} \sum_{j=1}^{M} (\hat{\boldsymbol{\epsilon}}_{\boldsymbol r}^{(j)},\hat{ \boldsymbol{\sigma}}_{\boldsymbol e}^{(j)}),
\end{equation}
where  $(\hat{\boldsymbol{\epsilon}}_{\boldsymbol r}^{(j)}, \hat{ \boldsymbol{\sigma}}_{\boldsymbol e}^{(j)}) \sim p(\boldsymbol{\epsilon_r}, \boldsymbol{\sigma_e} | \mathbf{d}_{obs})$  denotes the paired samples drawn from  $p(\boldsymbol{\epsilon_r}, \boldsymbol{\sigma_e} | \mathbf{d}_{obs})$,  indexed by $j$, with the number of all samples being $M$.

In the reminder of the paper, we are interested in obtaining these samples from the posterior  $p(\boldsymbol{\epsilon_r}, \boldsymbol{\sigma_e} | \mathbf{d}_{obs})$ proportional to
\begin{equation}\label{posterior_pixel}
	p(\boldsymbol{\epsilon_r}, \boldsymbol{\sigma_e} | \mathbf{d}_{obs}) \propto p(\mathbf{d}_{obs} | \boldsymbol{\epsilon_r}, \boldsymbol{\sigma_e}) \cdot p(\boldsymbol{\epsilon_r}, \boldsymbol{\sigma_e}),
\end{equation}
where $p(\mathbf{d}_{obs} | \boldsymbol{\epsilon_r}, \boldsymbol{\sigma_e})$ is the {likelihood} described by the forward model (\ref{original_forward_model}), and  $p(\boldsymbol{\epsilon_r}, \boldsymbol{\sigma_e})$ is the {prior distribution} encoding any known statistical structure or physical constraints of the material properties. A key challenge arises from the fact that $p(\boldsymbol{\epsilon_r}, \boldsymbol{\sigma_e})$ is not available in closed-form. Hence, we learn it from datasets. We next specifically address how to learn the prior and then perform posterior sampling using it. 

\section{\ac{ldpnp} framework}\label{SecIII}
This section presents the \ac{ldpnp} framework that samples the posterior with the learned prior, which is rooted in \ac{pdpnp} \cite{xu2024provably,wu2024principled} but allows faster convergence and improved  accuracy.

\begin{figure*}[!]
	\centering
	\includegraphics[width=180mm]{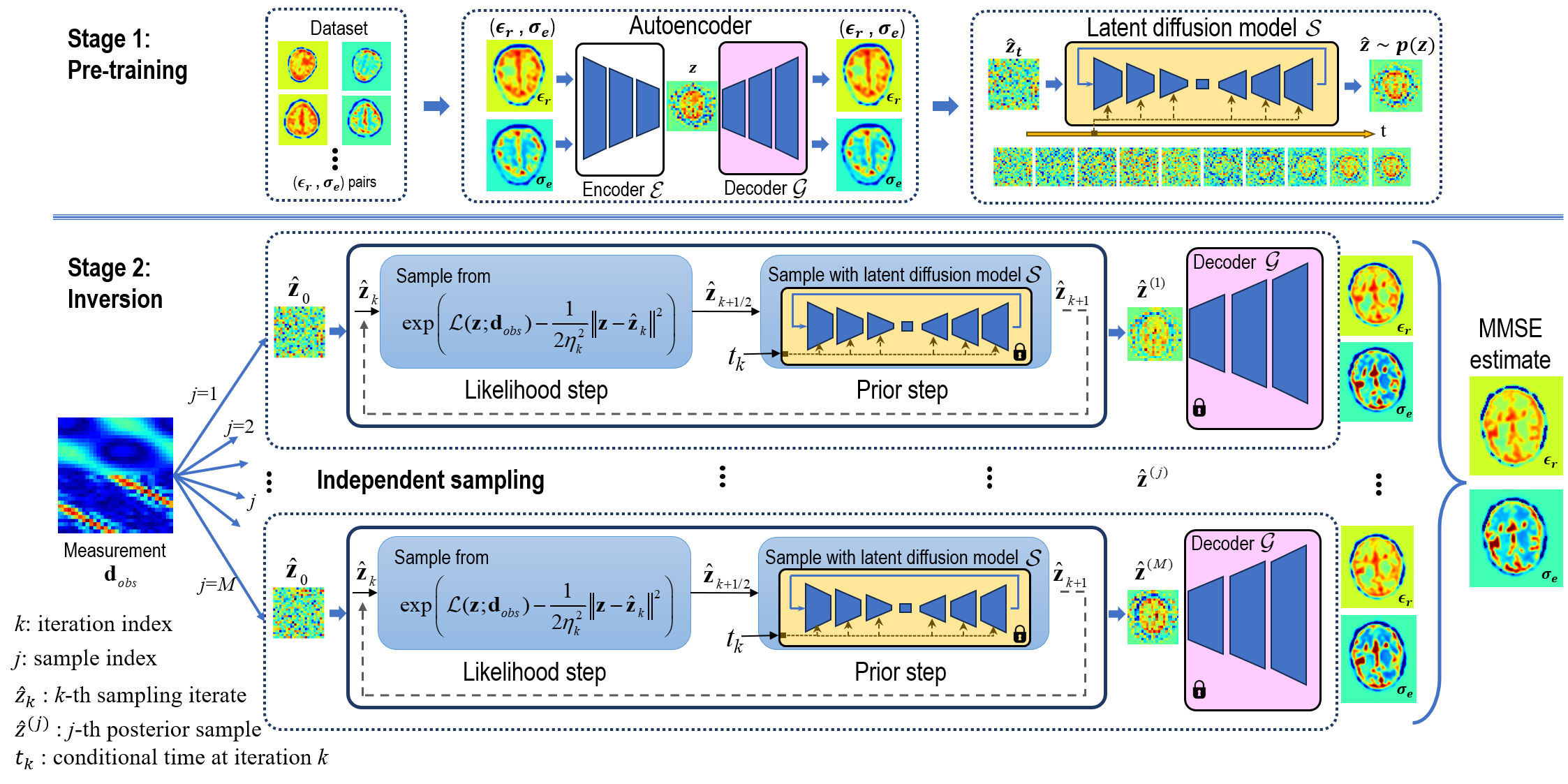}	
	\caption{Overview of the \ac{ldpnp} framework.  {Stage 1 (Pre-training):} An autoencoder is trained on paired \((\boldsymbol{\epsilon}_r, \boldsymbol{\sigma}_e)\) to obtain a latent representation \(\mathbf{z}\). A latent diffusion model is then trained on the \(\mathbf{z}\)-space to learn the prior distribution \(p(\mathbf{z})\). 
		{Stage 2 (Inversion):} Given observed measurement data \(\mathbf{d}_{{obs}}\), posterior samples in the latent space are iteratively drawn by alternating between a likelihood step (enforcing data consistency) and a prior step (guided by the diffusion model). The decoded samples are averaged to approximate the \ac{mmse} estimate of \((\boldsymbol{\epsilon}_r, \boldsymbol{\sigma}_e)\).}
	\label{fig_overview}
\end{figure*}

An overview of our framework is presented in Fig.~\ref{fig_overview}, which consists of a pre-training stage and an inversion stage. 

In the pre-training stage, given a dataset of paired samples $(\boldsymbol{\epsilon_r}, \boldsymbol{\sigma_e})$, 
we train an autoencoder to obtain an encoder $\mathcal E$ and a decoder $\mathcal G$. The encoder \(\mathcal{E}\) is used to map all \((\boldsymbol{\epsilon}_r, \boldsymbol{\sigma}_e)\) samples into corresponding latent representations \(\mathbf{z}\). Based on the collection of \(\mathbf{z}\) samples from the entire dataset,  a latent diffusion model $\mathcal S$ is trained, which can output samples from the prior distribution of $\mathbf z \sim p(\mathbf z)$. 

In the inversion stage, given observed measurement data \(\mathbf{d}_{{obs}}\), we independently draw  $M$ posterior samples in the latent space using an iterative sampling procedure. In each iteration, there is a likelihood step that promotes measurement fitness, and a prior step that encourages agreement with the learned prior. Specifically, the prior step is achieved by denoising the output of the likelihood step using the latent diffusion model $\mathcal S$. After obtaining all latent samples, we decode them using \(\mathcal{D}\) to reconstruct the corresponding \((\boldsymbol{\epsilon}_r, \boldsymbol{\sigma}_e)\) pairs. The \ac{mmse} estimate is approximated by the average of all $(\boldsymbol{\epsilon_r}, \boldsymbol{\sigma_e})$ pairs.

We begin by introducing how to jointly parameterize $\boldsymbol{\epsilon_r}$ and conductivity $\boldsymbol{\sigma_e}$ so that the unknowns jointly describe structures and materials. Based on this representation, the \ac{mmse} estimation problem is reformulated in the latent space. We then present an efficient method for estimating $\boldsymbol{\epsilon_r}$ and $\boldsymbol{\sigma_e}$ under this formulation. 

\subsection{Latent representation}
Representing $\boldsymbol{\epsilon_r}$ and $\boldsymbol{\sigma_e}$ with discrete image pixels results in high dimensionality, redundant details, varying sensitivity to observations, and loss of intrinsic material relationships. As we explain in Section \ref{latent_benefit},  this leads to high degrees of freedom and slow convergence in the solution. 

To overcome this issue, we suggest learning the parameterization from datasets, that is, projecting $\boldsymbol{\epsilon_r}$ and $\boldsymbol{\sigma_e}$ onto the latent space of an autoencoder. This approach allows to represent the joint distribution of $\boldsymbol{\epsilon_r}$ and $\boldsymbol{\sigma_e}$ in a compact low dimensional space, empowering the parameterization to have material-specific characteristics \cite{10332205}.

We train a convolutional autoencoder on a dataset of $(\boldsymbol{\epsilon_r}, \boldsymbol{\sigma_e})$ image pairs. The autoencoder contains an encoder and decoder, denoted as $\mathcal E$ and $\mathcal G$, respectively. The encoder learns to map the input $(\boldsymbol{\epsilon_r}, \boldsymbol{\sigma_e})$ image pair to a low-dimensional latent variable $\mathbf z$, that is, 
\begin{equation}\label{encoder}
	\mathbf z  = \mathcal E(\boldsymbol{\epsilon_r}, \boldsymbol{\sigma_e}) \in \mathbb{R}^{N_u \times N_v}
\end{equation} 
with $N_u < N_x$ and $N_v < N_y$. The decoder learns to reconstruct the original pair from $\mathbf z$, that is,
\begin{equation}\label{decoder}
		(\boldsymbol{\epsilon_r}, \boldsymbol{\sigma_e}) = \mathcal G(\mathbf z).
\end{equation}
The training of the autoencoder can be achieved by a self-supervised manner, that is, minimizing the reconstruction error between its output and input. The network structure and training details are given in  Appendix \ref{appendix_A} and  \ref{appendix_B}, respectively.

%

\subsection{Inverse problem in the latent space}
This subsection shows how the \ac{mmse} estimation (\ref{eq:mmse_joint}) is achieved with latent the representation. Based on  (\ref{mmse_average}) and (\ref{decoder}), 
\begin{equation}\label{mmse_average2}
	(\hat{\boldsymbol{\epsilon}}_{\boldsymbol r}, \hat{\boldsymbol{\sigma}}_{\boldsymbol e})_{\text{MMSE}} \approx \frac{1}{M} \sum_{j=1}^{M} \mathcal G( \hat{\mathbf z}^{(j)}),
\end{equation}
where $\hat{\mathbf z}^{(j)}$ is the corresponding latent variable describing  ($\hat{\boldsymbol{\epsilon}}_{\boldsymbol r}^{(j)},\hat{ \boldsymbol{\sigma}}_{\boldsymbol e}^{(j)}$). Next, we explain how to generate $\hat{\mathbf z}^{(j)}$  given measurement $\mathbf d_{obs}$.

The posterior of $\mathbf z$ given   $\mathbf d_{obs}$ is proportional to 
\begin{equation}\label{posterior_z_1}
	p(\mathbf z|\mathbf d_{obs})\propto	p(\mathbf d_{obs}|\mathbf z)  \cdot p(\mathbf z),
\end{equation}
where the prior $p(\mathbf z)$ is induced by $p(\boldsymbol{\epsilon_r}, \boldsymbol{\sigma_e})$ and the encoder $\mathcal E$ (\ref{encoder}). The likelihood $p(\mathbf d_{obs}|\mathbf z) $ is derived from the following forward process equivalent to our original forward model: 
\begin{equation}\label{key}
	\mathbf d_{obs} = F_{\mathcal G}(\mathbf z) + \mathbf n_{obs},
\end{equation}
where $F_{\mathcal G}(\cdot) = F(\mathcal G(\cdot)): \mathbb{R}^{N_u \times N_v} \rightarrow \mathbb{C}^{N_T N_R N_f}$ denotes a nonlinear map composed with the pre-trained decoder $\mathcal G$ (\ref{decoder}).  Consequently, assuming the  measurement noise is Gaussian with variance $\sigma^2$, the  likelihood $p(\mathbf d_{obs}|\mathbf z) $ is defined as 
\begin{equation}\label{likeli}
	p(\mathbf d_{obs}|\mathbf z) \propto  \exp (\mathcal L(\mathbf z; \mathbf d_{obs}) )  
\end{equation}
with
\begin{equation}\label{data_fidality}
	\mathcal L(\mathbf z; \mathbf d_{obs}) =  -\frac{1}{2\sigma^2} \| \mathbf d_{obs} - F_{\mathcal G}(\mathbf z)  \|^2.
\end{equation}
We can therefore transform our problem to sampling the latent variable $\hat{\mathbf z}^{(j)}$  from  the posterior (\ref{posterior_z_1}) in terms of $F_{\mathcal G}(\cdot)$ and $\mathbf z$,  after which the \ac{mmse} estimation (\ref{mmse_average}) are computed based on (\ref{mmse_average2}).

The choice of $p(\mathbf z)$ affects the algorithm used to estimate $\mathbf z$. For instance, when $p(\mathbf z)$ is uniform or Gaussian, the posterior $p(\mathbf z|\mathbf d_{obs})$ becomes Gaussian. This allows to solve $\mathbf z$ through deterministic optimization by maximizing the posterior, as studied in \cite{10332205}. 
However, in this work, we consider a more general setting where $p(\mathbf z)$ may have multiple peaks or be arbitrarily shaped, which makes closed-form or optimization-based solutions infeasible.

\subsection{Posterior sampling}\label{posterior_sampling}
We next focus on drawing samples from the posterior distribution in (\ref{posterior_z_1}) using the likelihood  (\ref{likeli}) and the prior \(p(\mathbf{z})\). This subsection assumes that $p(\mathbf z)$ is known; the method for obtaining it is provided in Section \ref{diffusion_into}. In the following, we describe the process for drawing a single sample and omit the superscript \((j)\) when there is no ambiguity. 

Standard  sampling methods, such as Metropolis-Hastings algorithm \cite{robert2004metropolis},  which requires a function proportional to the probability density,  or Langevin algorithm \cite{roberts1998optimal}, which relies on the log-density gradient, cannot be applied to sample from (\ref{posterior_z_1}), because we have no closed-form expression for its density, its gradient, or any equivalent unnormalized function. 

A result developed in \cite{Bouman} shows that a \ac{mc} constructed by alternately sampling from two proximal distributions—one derived from the prior and the other from the likelihood—converges to a stationary distribution that approximates the posterior. More specifically, the two proximal distributions of $p(\mathbf d_{obs}|\mathbf z)$ and $p(\mathbf z)$ are given by
\begin{equation}\label{q0_condition}
	q_{0}(\mathbf z|\mathbf v;\eta)\propto \exp \bigg(\mathcal L(\mathbf z; \mathbf d_{obs}) -\frac{1}{2\eta^2}  \| \mathbf z - \mathbf v\|^2 \bigg) , 
\end{equation}
\begin{equation}\label{q1_condition}
	q_{1}(\mathbf z|\mathbf v;\eta)\propto  p(\mathbf z)\exp\bigg(-\frac{1}{2\eta^2}  \| \mathbf z - \mathbf v\|^2\bigg) , 
\end{equation}
where $\mathbf v$ is a random vector  and $\eta$ is a parameter of the proximal distribution. When the new state of the \ac{mc} is generated by sampling (\ref{q0_condition}) and (\ref{q1_condition}) conditioned by the previous state, the distribution of the samples are close to the posterior as the chain is infinitely long and $\eta\xrightarrow{}0$. This result is important as it transforms the intractable task of posterior sampling into two simpler, tractable sampling problems.

Consequently, the samples of the \ac{mc} chain are obtained by an iterative approach. Given an iterate $\hat{\mathbf z}_k$ at the $k$-th iteration, a new sample $\hat{\mathbf z}_{k+1} $ is generated by sequentially drawing from two distributions:
\begin{enumerate}
	\item \textbf{Likelihood step:} 
	\begin{equation}\label{pcs}
		\hat{\mathbf z}_{k+\frac{1}{2}} \sim q_{0}(\mathbf z | \hat{\mathbf z}_{k};\eta_k),
	\end{equation}
	\item \textbf{Prior step:} 
	\begin{equation}\label{dds0}
		\hat{\mathbf z}_{k+1} \sim q_{1}(\mathbf z | \hat{\mathbf z}_{k+\frac{1}{2}};\eta_k),
	\end{equation}
\end{enumerate}
where $\eta_k$ is an annealing parameter in iterations. The likelihood step and prior step respectively promote consistency with the measurements and the prior knowledge. 

\begin{figure}[!]
	\centering
	\includegraphics[width=88mm]{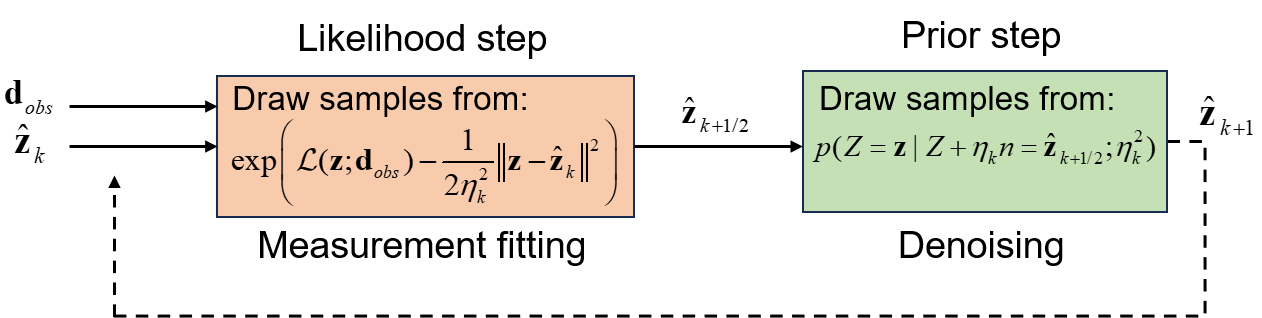}	
	\caption{One iteration of the iterative sampling approach. Given the iterate $\hat{\mathbf z}_k$ and the measurement $\mathbf d_{obs}$, the likelihood step samples an intermediate sample $\hat{\mathbf z}_{k+\frac{1}{2}}$ that promotes the measurement fitness. The prior step then generate a sample $\hat{\mathbf z}_{k+1}$ consistency with the prior, which  can be viewed as a denoising step.   }
	\label{fig_pcsdds}
\end{figure}

In Fig.~\ref{fig_pcsdds}, we show a block diagram of the iterative approach at each iteration,  explained in detail below. Given the measurement $\mathbf d_{obs}$ and the iterate $\hat{\mathbf z}_k$, the likelihood step  (\ref{pcs}) generates a sample $\hat{\mathbf z}_{k+\frac{1}{2}} $ that promotes measurement fitness $\mathcal L(\mathbf z; \mathbf d_{obs})$ (\ref{data_fidality}). This step can be viewed as sampling from an \textit{approximate} posterior distribution where the prior ${p(\mathbf z)}$ is replaced by a Gaussian distribution concentrated around the iterate $\hat{\mathbf z}_k$.  Since $q_{0}(\mathbf z | \hat{\mathbf z}_{k};\eta_k)$ is analytical, classical sampling algorithms can be employed to draw samples from it, as detailed in Section~\ref{numerical_sampling_likeli}.

After $\hat{\mathbf z}_{k+\frac{1}{2}}$ is obtained, the prior step (\ref{dds0}) draws $\hat{\mathbf z}_{k+{1}}\sim q_{1}(\mathbf z | \hat{\mathbf z}_{k+\frac{1}{2}};\eta_k)$. This step can be rigorously characterized by the denoising processing using a diffusion model  \cite{xu2024provably}. To see this, one can consider the second term of  (\ref{dds0}), $\exp\bigg(-\frac{1}{2\eta^2_k}  \| \mathbf z - \hat{\mathbf z}_{k+\frac{1}{2}}\|^2\bigg)$,  as a likelihood that is Gaussian,
\begin{equation}\label{gaussian_likeli}
	p(\hat{\mathbf z}_{k+\frac{1}{2}}|\mathbf z)\propto \exp\bigg(-\frac{1}{2\eta^2_k}  \| \hat{\mathbf z}_{k+\frac{1}{2}} - \mathbf z \|^2\bigg).
\end{equation}
Consequently, $ q_{1}(\mathbf z | \hat{\mathbf z}_{k+\frac{1}{2}};\eta_k)$ is proportional to 
\begin{equation}\label{dds}
	\begin{aligned}
		p(\mathbf z)  p(\hat{\mathbf z}_{k+\frac{1}{2}}|\mathbf z) \propto p(Z = \mathbf z| Z + \eta_k \mathbf n=\hat{\mathbf z}_{k+\frac{1}{2}}; \eta_k^2),
	\end{aligned}
\end{equation}
with $\mathbf n\sim\mathcal N(0,\mathbf I)$ where $\mathcal N$ denotes the Gaussian distribution. The expression (\ref{dds}), derived  from Bayes' rule, reformulates the sampling from (\ref{dds0}) as a denoising problem. Specifically, the input $\hat{\mathbf z}_{k+\frac{1}{2}}$ is interpreted as a noisy sample corrupted with Gaussian noise of variance $\eta_k^2$, and the prior step acts as a denoiser that takes this noisy input and produces a clean estimate. Its relationship with diffusion models will be further explained in Section \ref{diffusion_into} and  \ref{numerical_sampling_prior}.

We now summarize the  \ac{ldpnp} framework introduced so far. Given a dataset of paired $\boldsymbol{\epsilon_r}$ and $\boldsymbol{\sigma_e}$ maps, we first train an autoencoder to obtain an encoder $\mathcal E$ and a decoder $\mathcal G$, and a latent diffusion model describing $p(\mathbf z)$ (details in Section \ref{diffusion_into}). Next, we draw $M$ posterior samples in the latent space, that is, $\hat{\mathbf z}^{(j)}$ ($j=1,\dots, M$),  by repeatedly sampling. Each sample is obtained by running the likelihood step and prior step  for $N_k$ (pre-defined) iterations using \(\mathcal{G}\) and the measurement \(\mathbf{d}_{obs}\), as  detailed in  Section \ref{sec_implementation}. Finally, the estimates of permittivity and conductivity is computed as the average of the $M$ samples according to  (\ref{mmse_average2}). The entire  procedure is outlined in Algorithm~\ref{alg:ldpnp}.

\begin{algorithm}[!]
	\caption{\ac{ldpnp} framework}
	\label{alg:ldpnp}
	\begin{algorithmic}[1]
		\STATE \textbf{Require:} Dataset of  $(\boldsymbol{\epsilon_r}, \boldsymbol{\sigma_e})$ pairs.
		\STATE \textbf{Pre-training:} Train an autoencoder on $(\boldsymbol{\epsilon_r}, \boldsymbol{\sigma_e})$ pairs to obtain encoder $\mathcal{E}$ and decoder $\mathcal{G}$. Train a latent diffusion model $\mathcal S$ on all $\mathbf z=\mathcal E(\boldsymbol{\epsilon_r}, \boldsymbol{\sigma_e})$ maps. 
		\STATE \textbf{Input:}  Measurement data $\mathbf{d}_{obs}$,  number of samples $M$, number of iterations $N_k$.

		\vspace{0.5em}
		
		\FOR{$j = 1$ to $M$}
		\FOR{$k = 0$ to $N_k-1$}
		\STATE \textbf{Likelihood step} (\ref{pcs}):  	\[ \text{Draw}\quad  \hat{\mathbf{z}}_{k+\frac{1}{2}} \sim 	\exp \bigg(\mathcal L(\mathbf z; \mathbf d_{obs})-\frac{1}{2\eta^2_k}  \| \mathbf z - \hat{\mathbf z}_k\|^2 \bigg),\] 
	
		\STATE \textbf{Prior step}  (\ref{dds}):  \[ \text{Draw}  \quad \hat{\mathbf{z}}_{k+{1}} \sim  	p(Z = \mathbf z| Z + \eta_k n=\hat{\mathbf z}_{k+\frac{1}{2}}; \eta_k^2),\] 
		\ENDFOR
		\STATE Save final sample: $\hat{\mathbf{z}}^{(j)} = \hat{\mathbf{z}}_{N_k},$
		\ENDFOR
		
		\STATE \textbf{MMSE estimation:}  $	(\hat{\boldsymbol{\epsilon}}_{\boldsymbol r}, \hat{\boldsymbol{\sigma}}_{\boldsymbol e})_{\text{MMSE}} \approx \frac{1}{M} \sum_{j=1}^{M} \mathcal G( \hat{\mathbf z}^{(j)}),$
		\STATE \textbf{Output:} $	(\hat{\boldsymbol{\epsilon}}_{\boldsymbol r}, \hat{\boldsymbol{\sigma}}_{\boldsymbol e})_{\text{MMSE}}.$
	\end{algorithmic}
\end{algorithm}

\section{Implementation of \ac{ldpnp}}\label{sec_implementation}
This section presents the algorithm for drawing a single posterior sample, as summarized in Algorithm \ref{al1}. We begin by presenting the numerical sampling algorithms for the likelihood step.  Next, we introduce the learning of  \( p(\mathbf z) \) using a diffusion model, which provides the foundation for implementing the prior step. Finally, we establish the connection between the diffusion model and the prior step, and detail the sampling procedure.
\begin{algorithm}[t]
	\caption{\ac{ldpnp} posterior sampling (single sample)}
	\label{al1}
	\begin{algorithmic}[1] 
		\STATE \textbf{Require:} Pre-trained decoder  $\mathcal G(\cdot)$, pre-trained score model $\mathbf \mathcal S (\cdot, \cdot)$, forward model $F_{\mathcal G}(\cdot)$.
		\STATE \textbf{Input:} Measurements $\mathbf d_{obs}$, drift coefficient $f(t)$, diffusion coefficient $g(t)$, and transition probability variance $\beta^2(t)$.\hspace{-1mm}
		\STATE \textbf{Hyperparameters:}  The number of alternating the likelihood-prior step loops $N_k$, number of iteration $N_\tau$ in the likelihood step, number of iteration $N_t$ in the prior step,  annealing schedule $\eta_k$,  and the discrete time inteval $\gamma$ and $\delta$ in the likelihood step and the prior step.
		\vspace{0.5em}
		\STATE \textbf{Initialization:} $\hat{\mathbf z}_0$.
		\STATE \textbf{Sampling: }
		\FOR{$k = 0$ \textbf{to} $N_k-1$}    
		\vspace{0.5em}
		\STATE \textbf{Likelihood step:} Draw samples from (\ref{pcs}) :
		\STATE $\mathbf z[0] = \hat{\mathbf z}_k$,
		\STATE $r=e^{-\gamma/\eta_k^2}$,
		\FOR{$n = 0$ to $N_\tau-1$}
		\STATE $\mathbf n \sim \mathcal N(0,\mathbf I) $,
		\STATE $\displaystyle
		\begin{aligned}
			{\mathbf z}[n+1] &= \eta_k^2(1-r)\nabla_{{\mathbf z}[n]}\mathcal L(\mathbf z[n]; \mathbf d_{obs}) + r {\mathbf z}[n] \\
			&\quad + (1-r) \hat{\mathbf z}_k + \eta_k \sqrt{1-r^2}\mathbf n,
		\end{aligned}
		$ 
		\ENDFOR
		\STATE $\hat{\mathbf z}_{k+\frac{1}{2}} = {\mathbf z}[N_{\tau}]$.
		\vspace{0.7em}
		\STATE \textbf{Prior step:} Denoising with (\ref{dds}): 
		\STATE $t_{N_t} = \beta^{-1}(\eta_k)$,
		\STATE $\delta = {\frac{t_{N_t}-\varepsilon_t}{N_t}} $,
		\STATE $\mathbf z[N_t] = \hat{\mathbf z}_{k+\frac{1}{2}} $,
		\FOR{$n=N_t$ to $1$ }  
		\STATE $\mathbf n \sim \mathcal N(0,\mathbf I) $,
		\STATE $t_n = t_{N_t} - (N_t - n)\delta$,
		\STATE  $\displaystyle
		\begin{aligned}
			\mathbf z[n-1]& = \mathbf z[n] +\big ( g^2(t)\mathcal S (\mathbf z[n], t_n)- f(t_n)\mathbf z[n]\big )\delta \\& \quad +  g(t_n)\sqrt{\delta}\mathbf n,
		\end{aligned}
		$
		\ENDFOR
		\STATE $\hat{\mathbf z}_{k+1} = {\mathbf z}[0]$.
		\vspace{0.5em}        
		\ENDFOR
		\STATE \textbf{Output:} $\hat{\mathbf z} = \hat{\mathbf z}_{N_k}$.
	\end{algorithmic}
\end{algorithm}

\subsection{Numerical sampling in the likelihood step}\label{numerical_sampling_likeli}
The likelihood step is achieved using the Langevin dynamics algorithm \cite{roberts1998optimal}, which transforms the sampling problem into solving a \ac{sde}. Specifically, the \ac{sde} for sampling the likelihood (\ref{pcs}) is given by:
\begin{equation}\label{pcs_sde}
	\mathrm{d}\mathbf{z}_\tau = -\nabla \mathcal{L}(\mathbf{z}_\tau; \mathbf{d}_{\mathrm{obs}}) \, \mathrm{d}\tau + \frac{1}{\eta_k^2} (\mathbf{z}_\tau - \hat{\mathbf{z}}_k) \, \mathrm{d}\tau + \sqrt{2} \, \mathrm{d}\mathbf{w}_\tau,
\end{equation}
where $\tau$ is the time notation, and $\mathbf{w}_\tau$ represents a standard Wiener process.
Following \cite{xu2024provably}, we apply the exponential integrator to discretize (\ref{pcs_sde}) since the linear drift $\frac{1}{\eta_k^2}(\mathbf z_{\tau} - \hat{\mathbf z}_k )$ may be large. 
In the discrete domain, we represent the time as $\tau_{n} = n\gamma$, where $n = 0, \ldots, N_{\tau}$ with $\gamma$ being the time interval, and the corresponding discrete variable $\mathbf{z}_\tau$ is denoted as $\mathbf{z}[n]$. Based on (\ref{pcs_sde}), the sample at $\tau_{n+1} =(n+1)\gamma $ is computed by 
\begin{equation}\label{discrete_pcs}
	\begin{aligned}
		{\mathbf z}[n+1]&= \eta_k^2(1-r)\nabla_{{ {\mathbf z}}[n]}\mathcal L(\mathbf {\mathbf z}[n]; \mathbf d_{obs}) + r {\mathbf z}[n] \\& + (1-r) \hat{\mathbf z}_k  + \eta_k \sqrt{(1-r^2)}\mathbf n,
	\end{aligned}
\end{equation}
where $r=e^{-\gamma/\eta_k^2}$ and  $\mathbf n \sim \mathcal N(0,\mathbf I)$. 
We initialize $ \mathbf{z}[0] = \hat{\mathbf{z}}_k $, and after $ N_{\tau} $ iterations of the update, we obtain ${\mathbf z}[N_{\tau}]$. We treat this result as $\hat{\mathbf z}_{k+\frac{1}{2}}$, that is, $ \hat{\mathbf{z}}_{k+\frac{1}{2}} = \mathbf{z}[N_{\tau}] $.

\subsection{Diffusion models: preparation for the prior step}\label{diffusion_into}
In Section \ref{posterior_sampling}, we have seen that the prior step works as a denoiser that generates a clean sample given noisy input. This process can be rigorously described by the sampling (reverse) process of a diffusion  model \cite{xu2024provably}. We next introduce how a diffusion model learns the prior $p(\mathbf z)$.

Diffusion models estimate the prior distribution of a signal by training neural networks to reverse a gradual noising process. This technique achieves state-of-the-art performance in various generative tasks \cite{songscore, ho2020denoising,he2025diffusion,moser2024diffusion}. It can be modeled by a perturbation process and a reverse process. 

In the perturbation process, the datapoints are perturbed with a stochastic process over time $t\in [0,1]$, according to the following \ac{sde}
\begin{equation}\label{sde_fwd}
\text{d}\mathbf z_t = f(t) \mathbf z_t \text{d}t + g(t) \text{d}\mathbf w_t,
\end{equation}
where $\mathbf z_t$ is the trajectory of random variables in a stochastic process, $f(t)$ and $g(t)$ is the drift and diffusion coefficient of the \ac{sde}, respectively, and $\mathbf w_t$  is a standard Wiener process. At $t=0$, $ \mathbf z_0\equiv\mathbf z$, and distribution $p_0(\mathbf z)$ represents the prior $ p(\mathbf z)$: $p_0(\mathbf z)\equiv p(\mathbf z)$. The \ac{sde} gradually transforms $\mathbf z_0$ into noise, with transition distribution 
\begin{equation}\label{transition_p}
      p_{0t}(\mathbf z_t|\mathbf z_0) =\mathcal N(\mathbf z_t | \alpha(t)\mathbf z_0,\beta^2(t)\mathbf I).
\end{equation}
Here $\alpha(t)$ and $\beta(t)$ can be derived analytically from $f(t)$ and $g(t)$ \cite{sarkka2019applied}. 

The reverse process, used for sampling, transforms a noisy sample $\mathbf z_t $ back to a data sample $\mathbf z_0 \sim p_0(\mathbf z)$ by solving the reverse-time \ac{sde}:
\begin{equation}\label{reverse_sde}
    \text{d}\mathbf z_t = \big ( f(t)\mathbf z_t - g^2(t)\nabla_{\mathbf{z}_t}\log p_t(\mathbf z_t)\big)\text{d}t + g(t) \text{d}\bar{\mathbf w}_t,
\end{equation}
where \(\mathrm{d}t\) and \(\bar{\mathbf{w}}_t\) respectively denotes the time differential and a standard Wiener process running in reverse time. In particular, the score function denoted by $\nabla_{\mathbf{z}_t}\log p_t(\mathbf z_t)$ is approximated  training a time-dependent score model \(\mathcal{S}(\mathbf{z}_t, t)\) using denoising score matching \cite{vincent2011connection} on samples \(\{\mathbf{z}_0^{(i)}\}_{i=1}^{N_{\text{data}}} \sim p(\mathbf{z})\):
\begin{equation}\label{vae_loss}
\begin{aligned}
       & \min \frac{1}{N_{data}} \sum_{i=1}^{N_{data}} \mathbb{E}_{t \sim \mathcal{U}[0,1]} \mathbb{E}_{\mathbf z_t^{(i)} \sim p_{0t}(\mathbf z_t^{(i)}|\mathbf z_0^{(i)})} \\ &\left[ \|\mathcal S(\mathbf z_t^{(i)}, t) - \nabla_{\mathbf z_t^{(i)}} \log p_{0t}(\mathbf z_t^{(i)}|\mathbf z_0^{(i)}) \|_2^2 \right],
\end{aligned}
\end{equation}
where superscript \((i)\) indexes the training samples, \(\mathcal{U}[0,1]\) denotes the uniform distribution over the time interval \([0, 1]\), and \(p_{0t}(\mathbf{z}_t|\mathbf{z}_0)\) is the transition distribution defined in (\ref{transition_p}). In practice, the clean training datapoint $\mathbf z_0$ is generated by the encoder $\mathcal E$ (\ref{encoder}) and the noisy data $\mathbf z_t$ are generated according to the transition distribution (\ref{transition_p}). 


After training, we obtain \( \mathcal{S}(\mathbf{z}_t, t) \approx \nabla_{\mathbf{z}_t} \log p_t(\mathbf{z}_t) \). Plugging this into (\ref{reverse_sde}) and initializing the following \ac{sde} from $\mathbf{z}_T $ at time $T \in [0,1] $ enables the generation of samples consistent with $ p(\mathbf{z}) $:
\begin{equation}\label{reverse_sde_z}
	\text{d}\mathbf{z}_t = \big( f(t) \mathbf{z}_t - g^2(t) \mathcal{S}(\mathbf{z}_t, t) \big) \, \text{d}t + g(t) \, \text{d}\bar{\mathbf{w}}_t,
\end{equation}
where $ T $ and $\mathbf{z}_T $ denote the starting time and initial condition of the process, respectively.

\subsection{Numerical sampling in the prior step}\label{numerical_sampling_prior}
%
 
 We now establish the connection between (\ref{reverse_sde_z}) and the prior step (\ref{dds}). Recall that the prior step denoises the noisy input $ \hat{\mathbf{z}}_{k+\frac{1}{2}}$ with variance \( \eta_k^2 \) to produce a clean estimate. The likelihood of generating $ \hat{\mathbf{z}}_{k+\frac{1}{2}}$ from $ \hat{\mathbf{z}}_k $, given in (\ref{gaussian_likeli}), has the same form as the transition distribution (\ref{transition_p}). If we solve (\ref{reverse_sde_z}) with $ \mathbf{z}_T = \hat{\mathbf{z}}_{k+\frac{1}{2}}$, $ \beta(T) = \eta_k $, and $\alpha(t) \equiv 1$, where $ \alpha(t) $ and $ \beta(t) $ are determined by $ f(t) $ and $g(t) $, the solution process becomes equivalent to sampling from (\ref{dds}). An option of $ f(t) $ and $ g(t) $ is provided in Appendix~\ref{pdpnp_details}.  
 
The reverse-time \ac{sde}  (\ref{reverse_sde_z})  is numerically solved by the Euler-Maruyama approach \cite{kloeden1992stochastic}. We denote the sample at discrete time points $t_n=n\delta+ \varepsilon_t$  as $\mathbf z[n]$, where $n=N_t,\dots,0$, $\delta$ is the discrete time interval, and $\varepsilon_t$ is a small positive number. By using the finite difference, the sample at $t_{n-1}=(n-1)\delta+ \varepsilon_t$ is computed by
\begin{equation}\label{discrete_dds}
	\begin{aligned}
    \mathbf z[n-1]& = \mathbf z[n] +\big ( g^2(t_n)\mathcal S (\mathbf z[n], t_n)- f(t_n)\mathbf z[n]\big )\delta \\&+  g(t_n)\sqrt{\delta}\mathbf n,
	\end{aligned}
\end{equation}
where $\mathbf n \sim \mathcal N(0,\mathbf I)$. We initialize $\mathbf z[N_t] = \hat{\mathbf z}_{k+\frac{1}{2}}$ and $t_{N_t} = \beta^{-1}(\eta_k)$. After $N_t$ iterations of the update, we obtain $\mathbf z[0]$. The sample $\hat{\mathbf z}_{k+1}$ is taken as $\hat{\mathbf z}_{k+1} = \mathbf z[0]$.

\section{Experiments}\label{SecV}
\subsection{Datasets}
We test our method on three datasets that have diverging characteristics, derived from MNIST \cite{726791}, Fashion-MNIST \cite{xiao2017fashion}, and ATLAS \cite{liew2018large}. The first two datasets serve as algorithmic validation benchmarks \cite{li2018deepnis,wei2018deep,tom}. The measurement environment in these two scenarios resembles those in industrial non-destructive evaluation or radar-based imaging.

Reconstructing MNIST digits can be challenging in microwave because the cavities (hollow or enclosed regions) within the digits act as resonators, which leads to strong multiple scattering phenomena that increase the nonlinearity of the inverse problem.  Fashion-MNIST further extends the evaluation to inhomogeneous targets, allowing us to assess the model's ability to recover spatially varying materials.

Finally, we test our method on dataset derived from ATLAS (a clinical MRI dataset of anatomical brain images from stroke patients), which represents our target application in brain imaging. Accurate reconstruction in this setting is particularly challenging: brain tissues are highly inhomogeneous, exhibit strong attenuation of electromagnetic waves, and are enclosed by the skull, which reflects most of the incident energy. As a result, the scattered signals contain limited information about the interior, necessitating effective use of the measurement for high-fidelity reconstruction.

\subsubsection{Property map generation}
\textbf{MNIST} \& \textbf{Fashion-MNIST}: the original images of size 28 $\times$ 28 are upscaled to $N_x \times N_y = $ 64 $\times$ 64. The pixel values are normalized to the range 1–2, which corresponds to the relative permittivity of the targets. The size of domain $D$ is 0.30 m $\times$ 0.30 m. Conductivity is not considered in these two datasets, therefore the contrast matrix $\boldsymbol{\chi}$ is real-valued. The background relative permittivity is $\epsilon_{r,b}=1$, corresponding to the air. The number of training samples is 60K. 

\textbf{ATLAS}: The MRI images are segmented into five categories: skull, white matter, gray matter, cerebrospinal fluid (CSF), and the remaining parts including background and other tissues \cite{10332205}. We synthesize strokes using morphological irregularities with different electrical properties. We employ data augmentation, including scaling, rotation, flipping, and translation, to simulate a broader diversity of head shapes and poses. Based on \cite{10332205} and \cite{hamidipour2018electromagnetic}, we assign each tissue type and stroke with specific permittivity and conductivity values. Note that while the original image resolution in ATLAS is 1 mm$\times$ 1mm, we interpolate it onto a 3 mm $\times$ 3mm mesh, with $N_x \times N_y = $ 96 $\times$ 96, to accommodate the resolution limits of microwaves and ensure a reasonable runtime for \ac{em} simulations. The size of domain $D$ is 0.28 m $\times$ 0.28 m. The complex relative permittivity of the background medium is $\epsilon_{r,b}=44-17.9\text{j}$, corresponding to a matching medium. The number of training samples is 160K. 
\subsubsection{Synthetic measurement setup}
\textbf{MNIST} \& \textbf{Fashion-MNIST}: The number of transmitters and receivers is set to $N_T=16$, and $N_R=32$, respectively. The simulation frequency points are 1 GHz and 3 GHz. Both the transmitting and receiving antennas are uniformly distributed on a circle centered at the center of domain $D$, with a radius of 2 m. The incident fields are generated by the 2-D point source model. For each target, the simulated scattered field, denoted as$\mathbf {d^s}$, are corrupted with $nl=4\%$ Gaussian noise to produce the synthetic observation data:
\begin{equation}\label{noise_level_def}
    \mathbf d_{obs} =\mathbf {d^s} + nl \cdot std(\text{Real}(\mathbf {d^s})) \cdot \mathbf n_1 + \text{j} \cdot std(\text{Imag}(\mathbf {d^s})) \cdot \mathbf n_2,
\end{equation}
where $std(\cdot)$ means computing the standard deviation, Real($\cdot$) and Imag($\cdot$) means taking the real and imaginary part, respectively, and $\mathbf n_1, \mathbf n_2 \sim \mathcal N(0,\mathbf I)$.

\textbf{ATLAS}: We set $N_T=16$, and $N_R=32$. The simulation frequency points are 0.6 GHz and 1 GHz. The transmitting and receiving antennas are uniformly distributed on the circle with a radius of 0.14 m.  Incident fields are generated by the 2-D point source model. For each brain slice, the synthetic observation data are corrupted with $nl=20\%$ Gaussian noise according to (\ref{noise_level_def}), corresponding to a reasonable noise level in real cases.

\begin{figure*}[t]
	\centering
	\includegraphics[width=180mm]{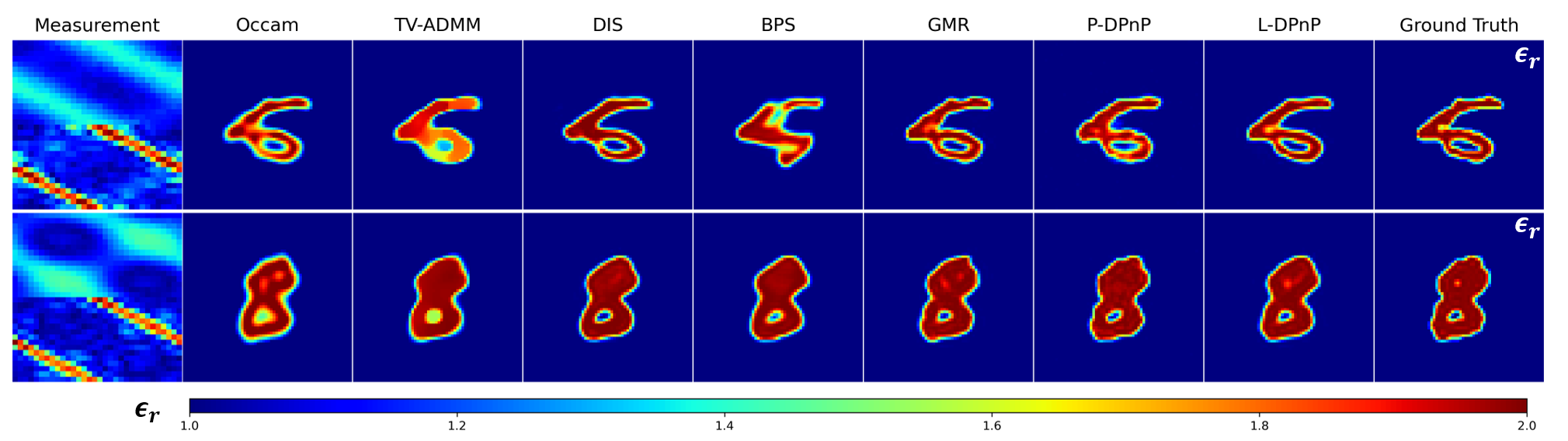}	
	\caption{ Results of permittivity ($\boldsymbol{\epsilon_r}$) reconstruction for MNIST targets. Only the amplitude of complex electrical fields are visualized in the measurement maps, with the upper half presenting the scattered field at 1 GHz, and the lower half presenting the scattered field at 3 GHz.  }
	\label{mnist_visual_comparison}
\end{figure*}

\begin{figure*}[t]
	\centering
	\includegraphics[width=180mm]{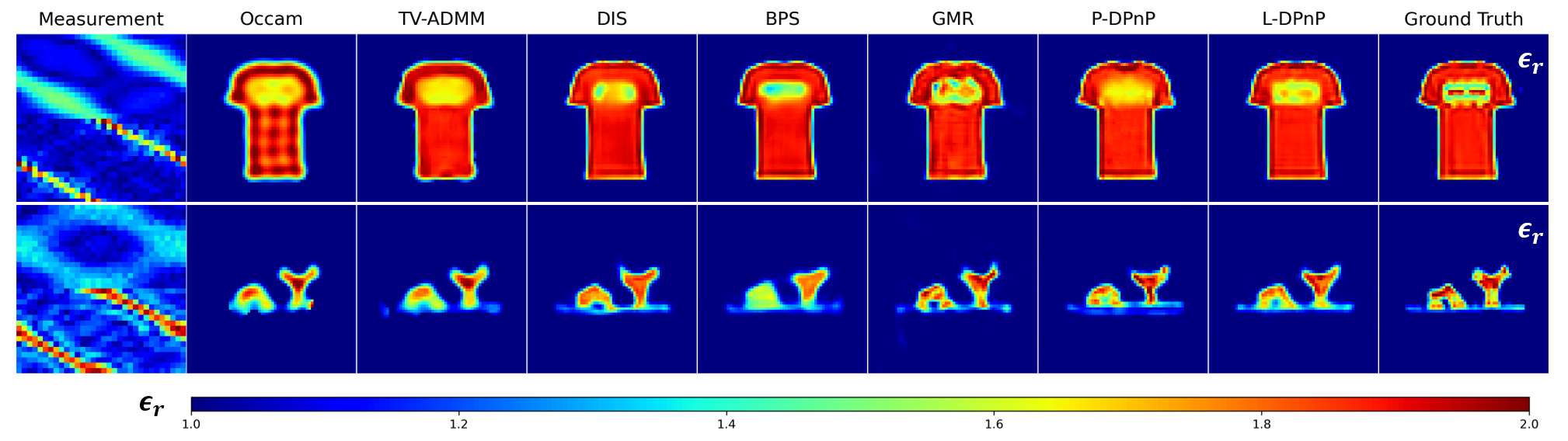}	
	\caption{ Results of permittivity ($\boldsymbol{\epsilon_r}$) reconstruction for Fashion-MNIST targets.  Only the amplitude of complex electrical fields are visualized in the measurement maps, with the upper half presenting the scattered field at 1 GHz, and the lower half presenting the scattered field at 3 GHz. }
	\label{fashion_mnist_visual_comparison}
\end{figure*}

\begin{figure*}[t]
	\centering
	\includegraphics[width=176mm]{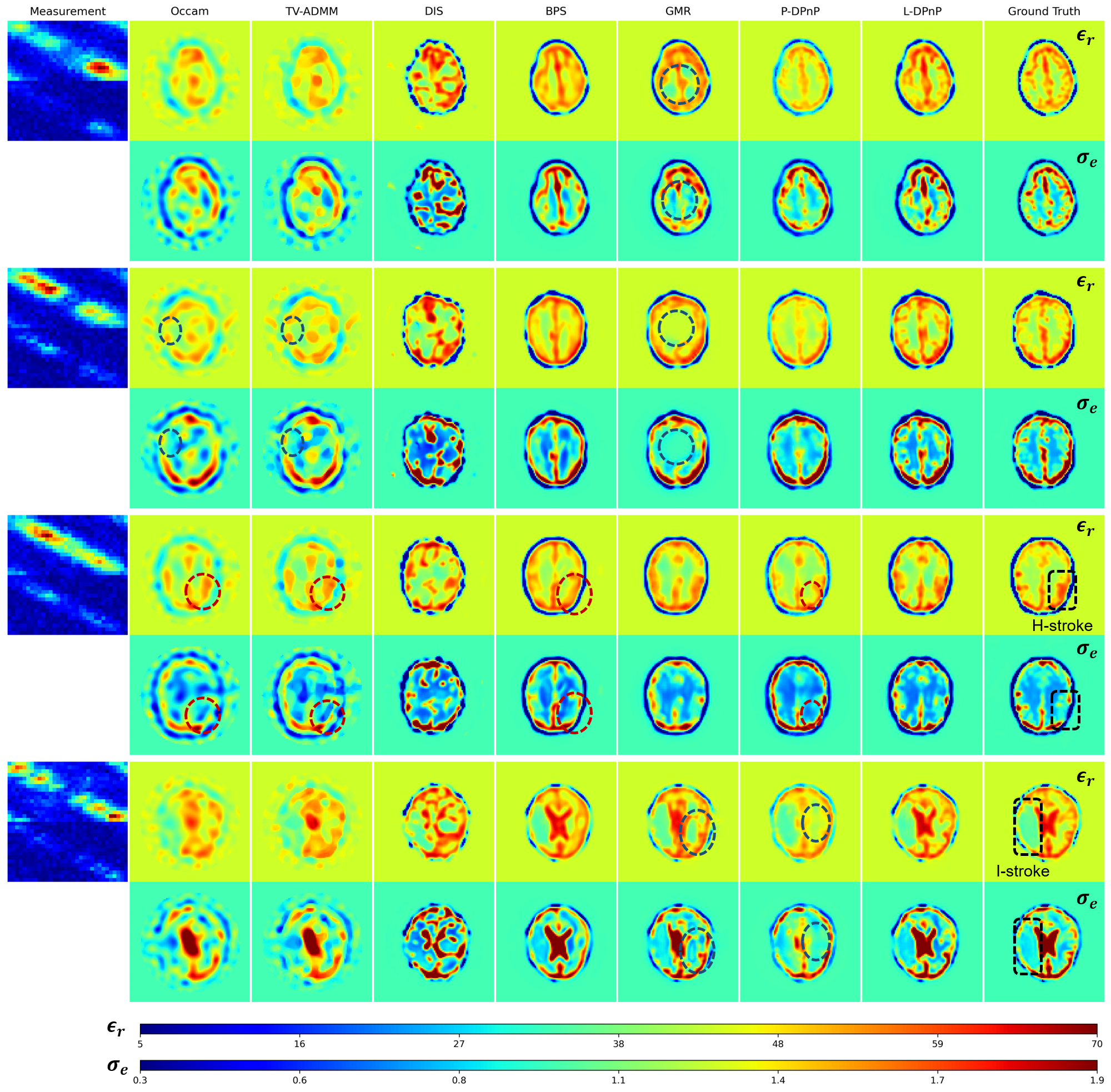}	
	\caption{ Results of permittivity ($\boldsymbol{\epsilon_r}$) and conductivity ($\boldsymbol{\sigma_e}$) reconstruction for brains. From top to bottom, two normal cases, one case with hemorrhagic stroke (h-stroke), and one case with ischemic stroke (i-stroke) are shown. The h-stroke (red color in $\boldsymbol{\epsilon_r}$, and turquoise color in $\boldsymbol{\sigma_e}$) is characterized by higher permittivity and conductivity than brain tissues. The i-stroke (turquoise color in both $\boldsymbol{\epsilon_r}$  and turquoise  $\boldsymbol{\sigma_e}$) is characterized by lower permittivity but higher conductivity than tissues.  DIS cannot reflect meaningful results under high noise level. In other comparison methods, we point out some false alarms in the normal cases (dark blue circles) and missed detection (dark red circles) .  }
	\label{brain_visual_comparison}
\end{figure*}

\subsection{Methods for comparison}
This subsection introduces other popular methods we use for comparison. They can be grouped into three categories, i.e., standard learning-free techniques, model-based deep learning, and learning-based \ac{pnp}. For each category, we provide two comparison baselines. All relevant neural network architectures and implementation details can be found in Appendix \ref{appendix_A} and \ref{appendix_B}, respectively.
\subsubsection{Standard learning-free techniques} The first baseline is an iterative optimization method that minimizes the data fidelity term together with an $\ell_2$ norm of the spatial gradient of permittivity and conductive maps. This approach, denoted as ``{Occam}'', was proposed in geophysical \ac{em} inversion to produce smooth reconstructions \cite{constable1987occam}.  The second baseline is the iterative reconstruction method that employs total variation (TV) regularization using \ac{admm}, denoted as  ``{TV-ADMM}'' \cite{aghamiry2019admm,liu2020alternating}.
\subsubsection{Model-based deep learning}
The first baseline, called ``DIS'', uses the difference between measured total fields and simulated total fields -- computed with a guessed permittivity map \cite{tom} -- as the input of a convolutional neural network to predict the true permittivity map. It is noted that when the initial guess is set to the background,  this approach is equivalent to the method proposed in \cite{8565987}. The second baseline, ``BPS'' \cite{wei2018deep}, feeds a post-processed map after back-projection \cite{chen2018computational} into a convolutional neural network to predict the permittivity map.
\subsubsection{Learning-based \ac{pnp}}
The first baseline is based on generative model reparameterization, namely ``GMR'', where the permittivity and conductivity map is projected to latent variables that are solved via deterministic optimization using gradient descent methods. The idea was first proposed in \cite{bora2017compressed} and applied to microwave imaging in \cite{10332205}. The second baseline is the application of the \ac{pdpnp} \cite{xu2024provably} to our problem. We mention that directly applying \ac{pdpnp} without masking highly sensitive regions may lead to failure due to the challenges discussed in Section \ref{latent_benefit}. The results presented in ATLAS examples are obtained by performing the likelihood step with an appropriate mask.

\subsection{Results}
\subsubsection{Visual results}
The inversion results of different algorithms tested on the three datasets are shown in Fig.~\ref{mnist_visual_comparison}, Fig.~\ref{fashion_mnist_visual_comparison}, and Fig.~\ref{brain_visual_comparison}, respectively. Due to high computational cost, we take the average of five samples, that is, $M=5$ in (\ref{mmse_average}), as the \ac{mmse} estimate. It can be seen that our algorithm can recover both lossless and lossy targets with the finest details and the least artifacts, for example, the enclosed regions in Fig.~\ref{mnist_visual_comparison}, the inhomogeneous material in Fig.~\ref{fashion_mnist_visual_comparison}, and the skull, brain tissues, and stroke regions in Fig.~\ref{brain_visual_comparison}.

We see that the learning-free methods cannot handle multiscale details: the images have either over-smooth boundaries or insufficient inhomogeneity. Purely data-driven methods tend to generate unrealistic structures particularly when the target has complex features, especially under noisy environment (see DIS for brain imaging). GMR may be trapped in local minima, as illustrated in the second  examples in Fig.~\ref{brain_visual_comparison}. \ac{pdpnp} tends to produce images with low saturation and less detailed structures, indicating less accurate reconstruction of the electrical properties. In comparison, our method yields reconstructions that most closely match the ground truth.  For brain imaging, Fig.~\ref{brain_visual_comparison} shows that other comparison methods may cause false alarms or fail to detect strokes while our approach reliably reconstruct property maps indicating the existence of strokes.

\begin{table*}[t]
    \caption{Quantitative evaluation of seven methods on three datasets (each 200 samples). \textbf{Bold}: best, \underline{underline}: second best.}
    \centering
    \newcolumntype{C}{>{\centering\arraybackslash}X}
\begin{tabularx}{\textwidth}{|c|c|*{7}{C|}} 
\hline
  &  & \multicolumn{7}{c|}{Method}  \\ 
\cline{3-9}
Dataset & Metric & Occam & TV-ADMM & DIS & BPS & GMR & P-DPnP & L-DPnP \\
\hline
& RMSE (Measurement)$\downarrow$ & \textbf{0.053} & 0.060 & 0.070 & 0.096 & \underline{0.055} & 0.058 & 0.059 \\
MNIST& RMSE (Reconstruction)$\downarrow$ & 0.040 & 0.053 & 0.033 & 0.046 & \underline{0.029} & 0.033 & \textbf{0.027} \\
& SSIM (Reconstruction)$\uparrow$ & 0.976 & 0.964 & 0.983 & 0.973 & \underline{0.984} & 0.983 & \textbf{0.987} \\\cline{1-9}
 & RMSE (Measurement) $\downarrow$&\underline{0.054} & \textbf{0.054} & 0.079 & 0.085 & 0.055 & 0.057 & 0.062 \\
Fashion-MNIST & RMSE (Reconstruction)$\downarrow$ & 0.056 & 0.053 & \underline{0.043} & 0.050 & 0.043 & 0.045 & \textbf{0.039} \\
& SSIM (Reconstruction)$\uparrow$& 0.854 & 0.881 & \underline{0.908} & 0.887 & 0.899 & 0.903 & \textbf{0.913} \\\cline{1-9}
 & RMSE (Measurement)$\downarrow$  & \underline{0.180} &\textbf{0.165} & 0.405 & 0.248 & 0.245 & 0.222 & 0.192 \\
 ATLAS& RMSE (Reconstruction)$\downarrow$ & 0.129 &0.128 & 0.156 & \underline{0.086} & 0.089 & 0.099 & \textbf{0.075} \\
 & SSIM (Reconstruction)$\uparrow$ & 0.695 & 0.711 & 0.814 & \underline{0.910} & 0.897 & 0.886 & \textbf{0.922}  \\
\hline
\end{tabularx}
    \label{tab:quantative_comparison}
\end{table*}

\begin{figure*}[t]
	\centering
	\includegraphics[width=180mm]{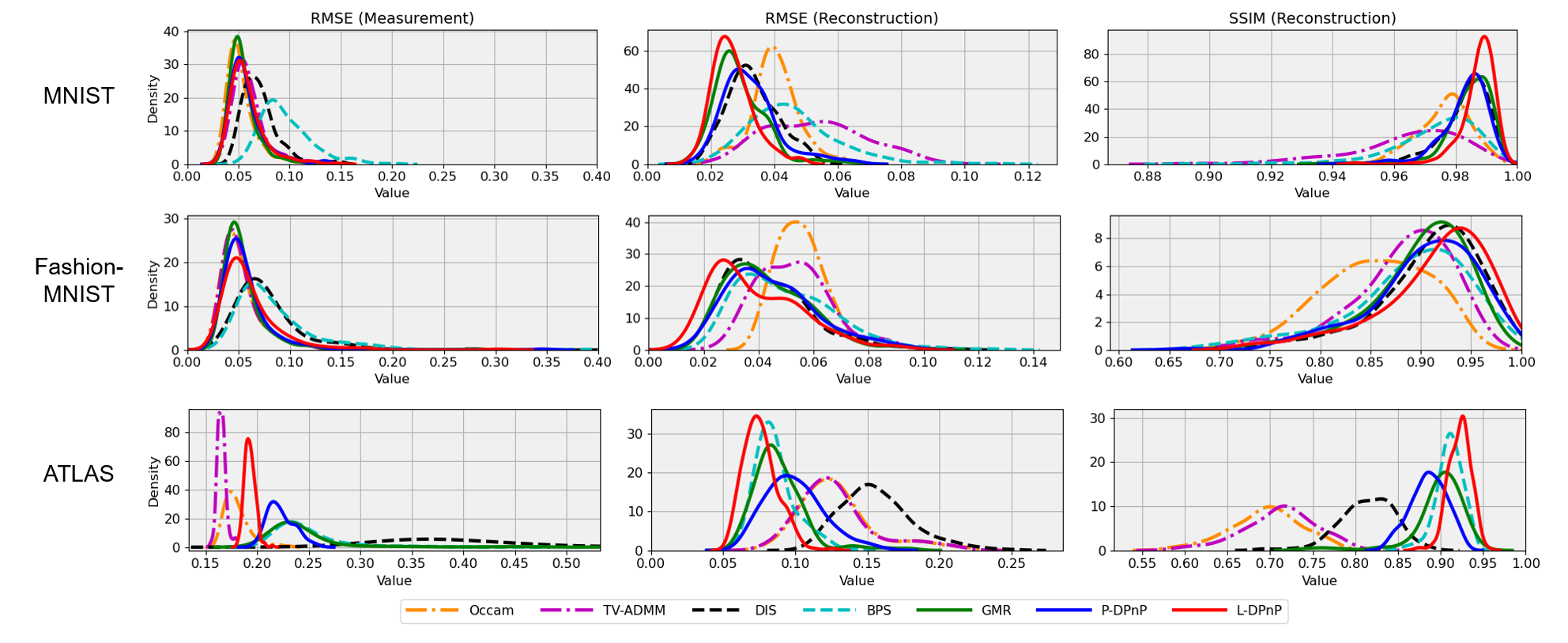}	
	\caption{Statistics of \ac{rmse} in the measurement domain, \ac{rmse} and \ac{ssim} in the reconstruction domain for different methods tested on the three datasets. }
	\label{evaluation_statistics}
\end{figure*}

\subsubsection{Performance metric} We report the performance using three metrics: \ac{rmse} in the measurement domain, \ac{rmse} in the reconstruction domain, and \ac{ssim} in the reconstruction domain. These metrics evaluate inversion quality from different yet complementary perspectives. \ac{rmse} in the measurement domain quantifies how well the simulated waves match the actual measurements, which is critical because a visually pleasing permittivity map might still be physically implausible if it fails to accurately reproduce the measurement data.  In practical detection scenarios, measurement fitness often serves as the primary quantitative criterion before the ground truth is precisely known. However, relying solely on \ac{rmse} in the measurement domain cannot fully assess imaging performance; for example, a low measurement-domain \ac{rmse} does not necessarily guarantee a meaningful reconstruction (as observed in learning-free methods). We further use \ac{rmse} in the reconstruction domain to quantify recovered electrical property that determines the material composition, and \ac{ssim} in the reconstruction domain to indicate how realistic of the reconstruction in terms of structures. Together, achieving  low \ac{rmse} in the measurement domain, along with low \ac{rmse} and high \ac{ssim} in the reconstruction domain, indicates a reliable reconstruction that is physically plausible, quantitatively accurate, and structurally realistic.

The comparisons of different methods are summarized in Tab.~\ref{tab:quantative_comparison}. We tested 200 cases for each dataset due to the high computational complexity of the problem. We observe that physics-driven, learning-free methods yield the lowest measurement \ac{rmse}, however, their reconstruction \ac{rmse} and \ac{ssim} are unsatisfactory due to the absence of prior knowledge.  In particular, when imaging the brain with high noises, the physics-driven methods achieve much lower  measurement \ac{rmse} than the preset noise level, which is a sign of overfitting to noise.  Model-based deep learning methods achieve improved reconstruction \ac{rmse} and \ac{ssim} compared to learning-free methods, but they exhibit higher measurement  \ac{rmse},  which raises concerns about reliability. \ac{pnp} methods offer a favorable balance between measurement and reconstruction fidelity. Specifically, our method achieves the highest reconstruction quality while maintaining measurement consistency near the noise level. The statistics of the metrics in Fig.~\ref{evaluation_statistics} show that our methods exhibit small \ac{rmse} and \ac{ssim} variance, indicating high inversion reliability.

\begin{figure}[t]
	\centering
	\includegraphics[width=88mm]{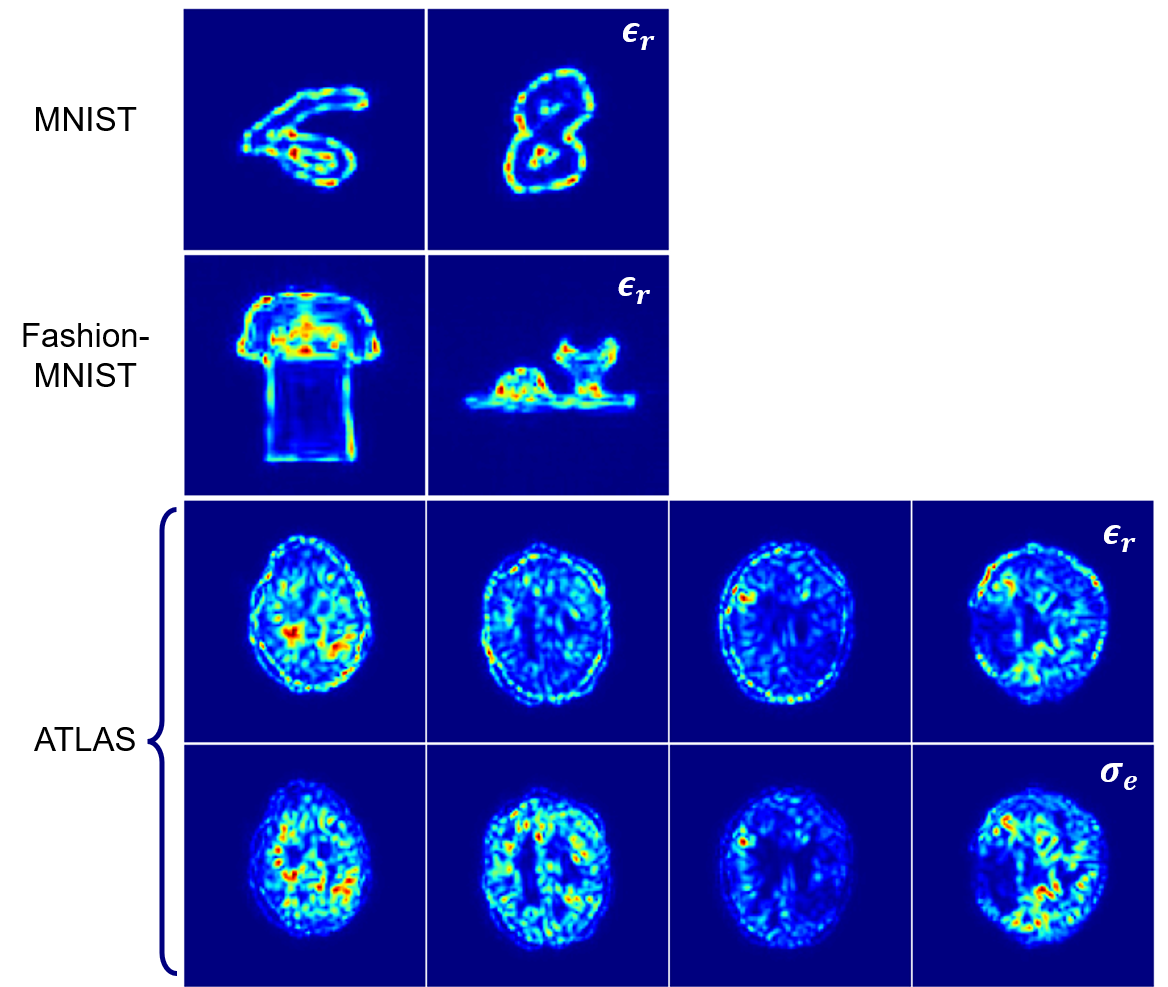}	
	\caption{Standard deviation of the reconstructed permittivity ($\boldsymbol{\epsilon_r}$) and conductivity ($\boldsymbol{\sigma_e}$) corresponding to  previous examples.   }
	\label{uncertainty}
\end{figure}

\subsubsection{Imaging uncertainty}Compared to deterministic optimization methods, sampling methods enable uncertainty quantification. As an additional demonstration, we present the standard deviation map corresponding to the reconstructed results from previous examples, see Fig.~\ref{uncertainty}. Regions with high standard deviation consistently correspond to areas exhibiting large reconstruction errors.

\section{Discussions}\label{sec:discussion}
\subsection{Benefits of the latent space}\label{latent_benefit}
\ac{ldpnp} offers three advantages over \ac{pdpnp}. First, by operating in the latent space rather than pixel space, the latent diffusion model achieves higher efficiency in both training and sampling \cite{rombach2022high}: eliminating imperceptible pixel‐level details reduces data dimensionality and accelerates convergence in training, while the lower-dimensional latent space enables faster sample generation.

Second, sampling in the latent space allows to generate more physically grounded targets than in the pixel space, which supports reliable interpretation. This is because the autoencoder jointly maps structural and material information into a coherent  latent representation.  By sampling latent codes within this manifold and decoding them, the permittivity and conductivity maps preserve realistic spatial–physical patterns. In contrast,  pixel-wise representation of permittivity and conductivity treats each image pixel independently, which increase the risk of producing physically implausible  material. 

\begin{figure}[t]
	\centering
	\includegraphics[width=88mm]{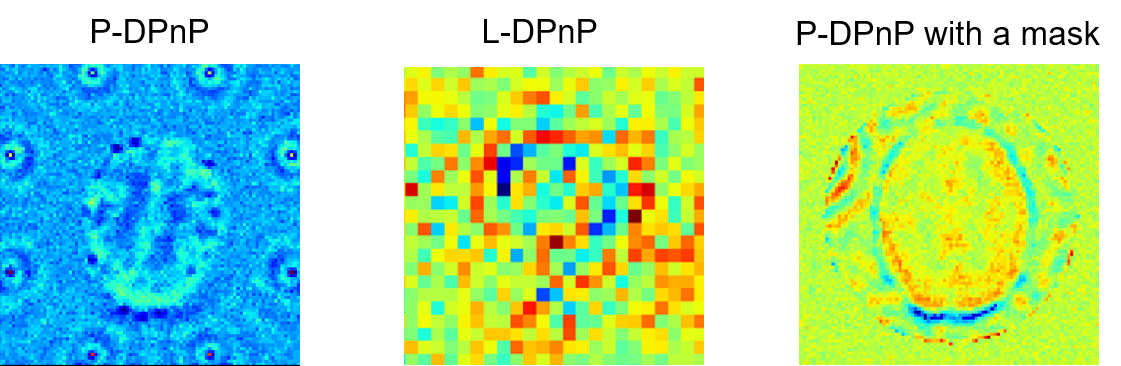}	
	\caption{{Output of the likelihood step using \ac{pdpnp} \cite{xu2024provably} and \ac{ldpnp}. The first map shows the output of the likelihood step in \ac{pdpnp}, which exhibits large values near high-sensitivity regions. The circular areas indicate the receivers' location.  The second map shows the output of the likelihood step in \ac{ldpnp}, which is less structured thanks to the rescale of sensitivity in the latent space. The third map shows that masking out the receivers  mitigates sensitivity variation, but is still less balanced and appears ripples due to the sensitivity difference of \ac{em} waves within a wavelength. }   }
	\label{fig1}
	\vspace{-0.4cm}
\end{figure}

Most importantly, \ac{ldpnp} overcomes slow convergence of \ac{pdpnp}, thanks to more scalable sensitivities in the latent space than in the pixel space. Recall that (\ref{discrete_pcs}) is computed based on the gradients  that indicate the sensitivity of unknown parameters to measurements. The pixel-based discretization, which is based on spatial coordinates, causes the sensitivity maps to have large variations, because (1) the wave amplitude decays exponentially with distance, (2) the wave’s phase and amplitude change at different rates within each wavelength, and (3) the wave attenuation differs spatially with the material's conductivity. Consequently, the output of likelihood step  in the pixel space, as shown in the first map of Fig.~\ref{fig1}, tends to exhibit large values near high-sensitivity regions. The \ac{pdpnp} algorithm preferentially updates high-sensitivity regions while low-sensitivity regions are updated more slowly. Considering the large computational cost of our forward problem, this slow-converging sampling process becomes impractical in real-world applications. In our experiments, we cannot get an acceptable sample within a reasonable running time (up to $N_k=200$, around 6 hours). 

In contrast, in \ac{ldpnp}, the encoder and decoder act like regularizers that normalize latent variable sensitivity by using fewer parameters to represent only essential features. This produces a less structured and more balanced gradient map, as shown in the second map of Fig.~\ref{fig1}. Empirically, the output of the likelihood step in latent space does not have large variances, therefore the latent variables can update at similar rates during sampling. It is relatively easy to adjust a proper stepsize in  the likelihood step to achieve fast convergence ($N_k=20$, around 25 minutes). 

To mitigate the effect of varying sensitivity, we apply the  mask on the gradient map to remove the high sensitivity regions near receivers in this paper. However, as shown in the third column of Fig.~\ref{fig1}, the result in the likelihood step is still less balanced due to varying sensitivity across pixels. This adjustment accelerates convergence compared to sampling without the mask, however, our experiments show that the reconstruction accuracy is still not as accurate as \ac{ldpnp} with the same $N_k$.
\subsection{Limitations and future work}
\ac{ldpnp} faces higher computational challenges than deterministic optimization such as GMR \cite{10332205}. The majority of the computational cost is incurred in the likelihood step, which requires a total of $N_k \times N_\tau \times M$ gradient evaluations to generate an \ac{mmse} estimate. For denser computational mesh, the time and memory demands can become prohibitive. To address this, future work should  explore distributed parallelization of forward solvers and the use of surrogate or reduced-order forward models to enable more efficient gradient computations. 


The prior step, while benefiting from diffusion‐based denoising, also incurs cost proportional to the number of diffusion timesteps. With recent advances in diffusion models, integrating fast sampling schemes such as DDIM \cite{song2020denoising} or DPM‐Solver \cite{lu2022dpm}  could further accelerate convergence. 
 
Extending \ac{ldpnp} to 3-D \ac{em} imaging \cite{10332205} is an important next step, as real‐world targets often exhibit complex 3-D structures that cannot be accurately captured by 2-D approximations. However, this  extension  remains challenging since it requires designing scalable network architectures for learning the prior and faster 3-D solvers that can handle the increased memory and sampling time.  Beyond \ac{em},  applying \ac{ldpnp} to  the \ac{isp} of other imaging modalities such as ultrasound or electrical impedance tomography \cite{zhang2019three, li2023quantitative} offers the opportunity of super-resolution and uncertainty analysis in these domains. Moreover, \ac{ldpnp} for multi-modality joint imaging \cite{ma2024recent,guo2024deep,song2020three}, where the latent diffusion model could fuse information from different physical properties, may yield more robust and complementary reconstructions in the future.

\section{Conclusion}\label{SecVI}
We propose a new posterior sampling approach to solve highly nonlinear and ill-posed \ac{em} \ac{isp} based on latent diffusion. This approach contains a the likelihood step that promotes measurement fitness and a prior step that enforces prior knowledge in the latent space. It does not require paired training data and can flexibly adapt to heterogeneous measurement setups.  The framework is modular, which enables the flexible integration of physical models into the generative models to leverage prior knowledge to improve reconstruction performance. The samples yield low quantitative error and high-fidelity structures. After sampling, our approach enables \ac{mmse} estimation and uncertainty quantification, which improves the reliability of \ac{em} imaging. Experimental results demonstrate that our approach achieves state-of-the-art performance in terms of reconstruction fitness and structural similarity with high measurement fidelity, outperforming existing physics-driven and learning-based methods.  

\appendix  \label{appendix}
For notational consistency, we use $\mathbf X$ and $\mathbf X_{out} $ to denote the input and output of neural networks that operate in the pixel-based imaging (including Occam, TV-ADMM, DIS, BPS, \ac{pdpnp}), and $\mathbf Z$ and $\mathbf Z_{out} $  to denote the input and output of neural networks operating in the latent space (including GMR and \ac{dpnp}). The exact physical meanings and construction details of these inputs and outputs in different neural networks are provided in Appendix~\ref{appendix_B}. 
\subsection{Neural network architectures}\label{appendix_A}
Notations for basic neural network operation:
\begin{itemize}
	\item $\text{Conv2D}_{C_{{out}}}$: 2D convolution with a kernel size of $3 \times 3$, stride 1, and $C_{\text{out}}$ output channels.
	\item $\text{Conv2D}_{C_{{out}}, /2}$: 2D convolution with stride 2 for downsampling, producing $C_{\text{out}}$ output channels.
	\item $\text{Conv2D}_{C_{{out}}, \times 2}$: First applies upsampling with a scale factor of 2, followed by 2D convolution producing $C_{{out}}$ output channels.
	\item $\text{GroupNorm}_{C_{{out}}/2}$: Group normalization with $C_{{out}}/2$ groups.
	\item SiLU: Sigmoid Linear Unit activation function.
\end{itemize}

\subsubsection{Autoencoder for latent representation} \label{Autoencoder for latent representation}
\begin{figure}[!]
	\centering
	\includegraphics[width=88mm]{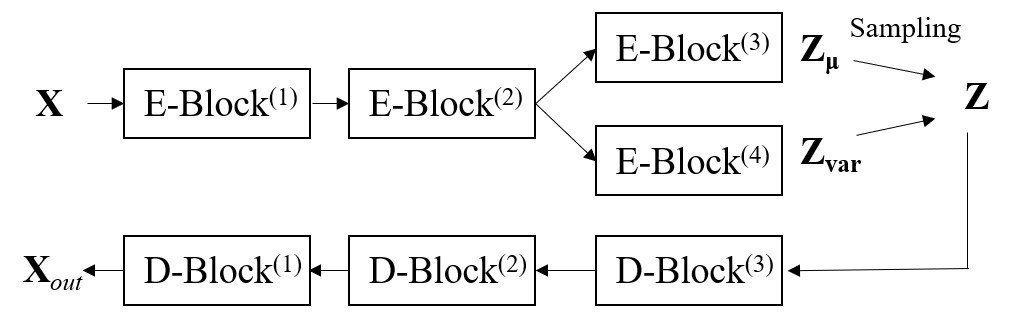}	
	\caption{Diagram of the autoencoder's architecture.    }
	\label{vae_architecture}
\end{figure}

The architecture's diagram is shown in Fig.~\ref{vae_architecture}, which is modified from \cite{10332205}. We apply a \ac{vae} as the autoencoder, suggested by \cite{rombach2022high}.  The encoding and decoding blocks are built with fully convolutional layers. Different from \cite{10332205}, where latent variables are 1D vectors, our latent map is 2D to adapt to latent diffusion. Denote the input and output tensor as $\mathbf X, \mathbf X_{out} \in\mathbb R^{B\times C\times H\times W}$, respectively, where $B$, $C$, $H$, and $W$ is the batch size, input channel number, height and width, respectively.  A mean tensor $\mathbf Z_{\mu}$ and logarithmic variance tensor $\mathbf Z_{var}$ is obtained by going through cascaded encoding blocks $\text{E-Block}^{(i)}_{C_{in}\xrightarrow{}C_{out}}$, where $C_{in}$ and $C_{out}$ represents the number of input and output channels of the block. Following the reparameterization trick \cite{kingma2013auto} of training a \ac{vae}, a latent sample $\mathbf Z$ is generated according to the mean and variance tensor. Then $\mathbf Z$ is decoded back by cascaded decoding blocks $\text{D-Block}^{(i)}_{C_{in}\xrightarrow{}C_{out}}$, yielding the output $\mathbf X_{out}$. This process is summarized in Architecture \ref{al:autoencoder_for_latent_1}.

As suggested by \cite{rombach2022high}, we utilize less aggressive downsampling in the autoencoder to attain a high decoding resolution. While \cite{rombach2022high} applied the vector quantized \ac{vae} (VQ-VAE) \cite{razavi2019generating} to avoid posterior collapse, our \ac{vae} adopts a small penalty on the KL divergence of the evidence lower bound (ELBO) \cite{kingma2013auto,pu2016variational} to address this problem. This autoencoder is expected to generate high-resolution images and have a more complete latent space than a classical autoencoder. After it is trained, the encoder $\mathcal E$ is used to generate datasets for training the score model, and the decoder is integrated into the forward model (\ref{original_forward_model}) to form $F_{\mathcal G}(\cdot)$.

\begin{Architecture}
	\caption{Autoencoder for latent representation}
	\label{al:autoencoder_for_latent_1}
	\begin{algorithmic}[1]  
		\STATE \textbf{Input:} $\mathbf X \in\mathbb R^{B\times C\times H\times W}$.
        \STATE $\mathbf X_{1} = \text{E-block}^{(2)}_{16\xrightarrow{} 64}(  \text{E-block}^{(1)}_{C\xrightarrow{} 16}(\mathbf X)),\nonumber$
        \STATE $\mathbf Z_{\mu} = \text{E-block}^{(3)}_{64\xrightarrow{} 1}(\mathbf X_1),\nonumber$
        \STATE $ \mathbf Z_{var} = \text{E-block}^{(4)}_{64\xrightarrow{} 1}(\mathbf X_1),\nonumber$
        \STATE $\mathbf n\sim\mathcal(0,\mathbf I)$,
        \STATE $ \mathbf Z = \mathbf Z_{\mu} + \exp (0.5 \mathbf Z_{var})  \cdot \mathbf n,\nonumber$
        \STATE $ \mathbf X_2 = \text{D-Block}^{(3)}_{1\xrightarrow{}64}((\mathbf Z)),\nonumber$
        \STATE $ \mathbf X_{out} = \text{D-block}^{(1)}_{16\xrightarrow{} C}(  \text{D-block}^{(2)}_{64\xrightarrow{} 16}(\mathbf X_2)).\nonumber$
 	\STATE \textbf{Output: $ \mathbf X_{out} \in\mathbb R^{B\times C\times H\times W}$}.
	\end{algorithmic}
\end{Architecture}

The $\big \{\text{E-Block}^{(i)}_{C_{in}\xrightarrow{}C_{out}}\big\}_{i=1,2} $ ($C_{out}=16$ for $i=1$ and $C_{out}=64$ for $i=2$) are similarly constructed. Denote the input and output of the blocks as $\mathbf X^{block}_{in}$ and $\mathbf X^{block}_{out}$,  respectively\footnote{The intermediate variable symbols are local to each network block. Identical symbols in different blocks do not refer to the same variable and should be interpreted according to context. }. Their structures are presented in Architecture \ref{al:e-block-12}.
Similarly, architectures of $\big \{\text{E-Block}^{(i)}_{64\xrightarrow{}1}\big\}_{i=3,4} $ are shown in Architecture \ref{al:e-block-mu-var}.

\begin{Architecture}
	\caption{$\big \{\text{E-Block}^{(i)}_{C_{in}\xrightarrow{}C_{out}}\big\}_{i=1,2} $}
	\label{al:e-block-12}
	\begin{algorithmic}[1]  
		\STATE \textbf{Input:} $\mathbf X^{block}_{in} \in\mathbb R^{B\times C_{in} \times H\times W}$.
        \STATE $    \mathbf X_1 = \text{SiLU}(\text{GroupNorm}_{C_{out}/2} (\text{Conv2D}_{C_{out}}(\mathbf X^{block}_{in}))), \nonumber $
        \STATE $ \mathbf X_2 = \text{SiLU}(\text{GroupNorm}_{C_{out}/2} (\text{Conv2D}_{C_{out}}( \mathbf X_1)), \nonumber$
        \STATE $ \mathbf X_2 = \text{SiLU}(\text{GroupNorm}_{C_{out}/2} (\text{Conv2D}_{C_{out}}( \mathbf X_2)), \nonumber$
        \STATE $\mathbf X_1^{\text{res}} =  \mathbf X_1 + 0.1 \mathbf X_2, \nonumber$
        \STATE $  \mathbf D_2 =  \text{SiLU}(\text{Conv2D}_{C_{out}}( \text{SiLU}(\text{Conv2D}_{C_{out}}(\mathbf X_1^{\text{res}})) )) , \nonumber$
        \STATE $  \mathbf D_2 =  \text{Conv2D}_{C_{out, /2}}(  \mathbf D_2  ) , \nonumber$
        \STATE $  \mathbf X^{block}_{out} =\mathbf D_1 + 0.1 \mathbf D_2. \nonumber$
 	\STATE \textbf{Output: $ \mathbf X^{block}_{out} \in\mathbb R^{B\times C_{out}\times H/2\times W/2}$}.
	\end{algorithmic}
\end{Architecture}

\begin{Architecture}
	\caption{$\big \{\text{E-Block}^{(i)}_{64\xrightarrow{}1}\big\}_{i=3,4}$}
	\label{al:e-block-mu-var}
	\begin{algorithmic}[1]  
		\STATE \textbf{Input:} $\mathbf X^{block}_{in} \in\mathbb R^{B\times C_{in} \times H\times W}$.
        \STATE $         \mathbf X_1 = \text{SiLU}((\text{Conv2D}_{32}(\mathbf X^{block}_{in}))), \nonumber
 $
        \STATE $ \mathbf X^{block}_{out} = \text{Conv2D}_{1}(\text{SiLU}((\text{Conv2D}_{16}(\mathbf X_1)))). \nonumber$
     \STATE \textbf{Output: $ \mathbf X^{block}_{out} \in\mathbb R^{B\times C_{out}\times H\times W}$}.
	\end{algorithmic}
\end{Architecture}

The architectures of the decoder blocks $ \text{D-Block}^{(3)}_{1\xrightarrow{}64}$ and $\big \{\text{D-Block}^{(i)}_{C_{in}\xrightarrow{}C_{out}}\big\}_{i=1,2} $ ($C_{out}=C$ for $i=1$ and $C_{out}=16$ for $i=2$) are presented in Architecture \ref{al:d-block-mu-var} and \ref{al:d-block-12}, respectively. 

\begin{Architecture}
	\caption{$\text{D-Block}^{(3)}_{1\xrightarrow{}64} $}
	\label{al:d-block-mu-var}
	\begin{algorithmic}[1]  
		\STATE \textbf{Input:} $\mathbf X^{block}_{in} \in\mathbb R^{B\times 1 \times H\times W}$
        \STATE $  \mathbf X_1 = \text{SiLU}((\text{Conv2D}_{16}(\mathbf X^{block}_{in}))), \nonumber$
        \STATE $ \mathbf X^{block}_{out} = \text{Conv2D}_{64}(\text{SiLU}((\text{Conv2D}_{32}(\mathbf X_1)))). \nonumber $
     \STATE \textbf{Output: $ \mathbf X^{block}_{out} \in\mathbb R^{B\times64\times H\times W}$}.
	\end{algorithmic}
\end{Architecture}

\begin{Architecture}
	\caption{$\big \{\text{D-Block}^{(i)}_{C_{in}\xrightarrow{}C_{out}}\big\}_{i=1,2} $}
	\label{al:d-block-12}
	\begin{algorithmic}[1]  
  
		\STATE \textbf{Input:} $\mathbf X^{block}_{in} \in\mathbb R^{B\times C_{in} \times H\times W}$.
        \STATE $ \mathbf X_1 = \text{SiLU}(\text{GroupNorm}_{C_{out}} (\text{Conv2D}_{2C_{out}}(\mathbf X^{block}_{in}))), \nonumber$
        \STATE $  \mathbf X_2 = \text{SiLU}(\text{GroupNorm}_{C_{out}} (\text{Conv2D}_{2C_{out}}(\mathbf X_2))), \nonumber$
        \STATE $    \mathbf X_2 = \text{SiLU}(\text{GroupNorm}_{C_{out}} (\text{Conv2D}_{2C_{out}}(\mathbf X_2))), \nonumber$
        \STATE $ \mathbf X_1^{\text{res}} =  \mathbf X_1 + 0.1 \mathbf X_2 , \nonumber$
        \STATE $  \mathbf U_1 = \text{Conv2D}_{C_{out}, \times 2}(  \mathbf X_1^{\text{res}}),$
        \STATE $  \mathbf U_2 =  \text{SiLU}(\text{Conv2D}_{C_{out}}( \text{SiLU}(\text{Conv2D}_{C_{out},\times 2}(\mathbf X_1^{\text{res}})))),$\STATE $ \mathbf U_2 =  \text{Conv2D}_{C_{out}}(  \mathbf U_2  ),$
        \STATE $ \mathbf X^{block}_{out} =\mathbf U_1 + 0.1 \mathbf U_2. \nonumber$
     \STATE \textbf{Output: $ \mathbf X^{block}_{out} \in\mathbb R^{B\times C_{out}\times 2H\times 2W}$}.
	\end{algorithmic}
\end{Architecture}

\subsubsection{Score model for \ac{ldpnp}} \label{Score model for ldpnp}
The latent diffusion model's architecture is based on U-Net. Time embedding is used to handle the temporal dynamics of the diffusion process. After it is trained, we obtain the score model $\mathcal S$ in (\ref{reverse_sde_z}).

\begin{figure}[!t]
	\centering
	\includegraphics[width=88mm]{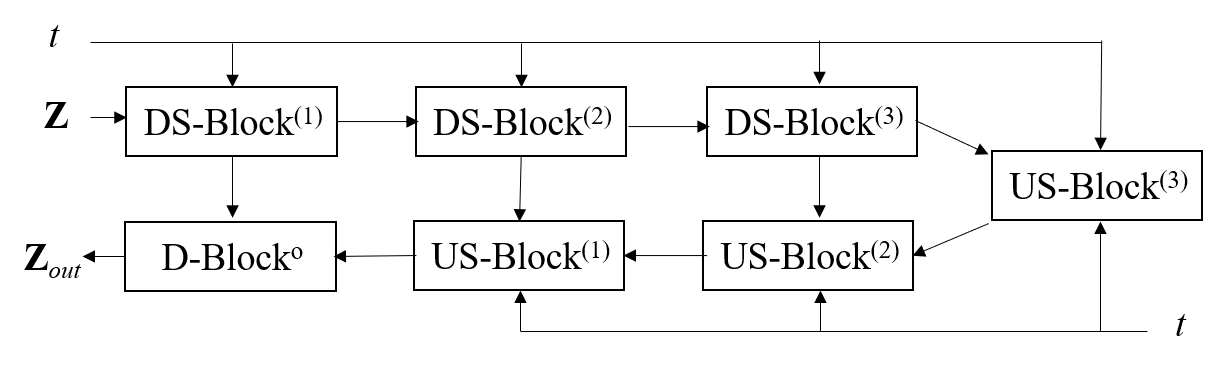}	
	\caption{Diagram of the score model's architecture.    }
	\label{score_architecture}
\end{figure}

Denote the input and output tensor as $\mathbf Z, \mathbf Z_{out} \in\mathbb R^{B\times 1 \times H\times W}$, where $B$,  $H$, and $W$ is the batch size,  height and width, respectively. The modified U-net is composed of cascaded down-sampling blocks $\text{DS-Block}_{C_{in}\xrightarrow{}C_{out}}^{(i)}$, up-sampling blocks $\text{US-Block}^{(i)}_{C_{in}\xrightarrow{}C_{out}}$, and output decoding block $ \text{D-Block}^{o}_{32\xrightarrow{}1}$, conditioned by the time-embedding vector $\mathbf{V}(t)\in\mathbb R^{B\times 512}$. The architecture's diagram is shown in Fig.~\ref{score_architecture}, with details in Architecture \ref{al:score_ldpnp}. 

In Architecture \ref{al:score_ldpnp}, time embedding is achieved by Gaussian Fourier projection. Given a batch of time values $t\in\mathbb R^B$, the projection generates high-dimensional features $\mathbf V_p(t)\in\mathbb R^{B\times 512}$ using fixed (unlearnable) random frequencies $\mathbf W\in \mathbb R^{256}$. $t\mathbf W$ is the element-wise multiplication (broadcasted across the batch dimension). The final time embedding $\mathbf V(t)$ is obtained after a fully connected layer $\text{Dense}_{512\xrightarrow{}512}$, whose input and output dimension are both 512.
\begin{Architecture}
	\caption{Score model for \ac{ldpnp}}
	\label{al:score_ldpnp}
	\begin{algorithmic}[1]  
		\STATE \textbf{Input:} $\mathbf Z \in\mathbb R^{B\times 1\times H\times W}$, $t\in\mathbb R^B$.
        \STATE $ \mathbf V_p(t) = \text{concat}[\sin(2\pi t\mathbf W),\cos(2\pi t\mathbf W) ],$
        \STATE $    \mathbf V(t)=\text{SiLU}(\text{Dense}_{512\xrightarrow{}512} (\mathbf V_p(t))), \nonumber$
        \STATE $  \mathbf Z_1^{skip}, \mathbf Z_1 = \text{DS-Block}_{C\xrightarrow{}32}^{(1)}(\mathbf Z,  \mathbf{V}(t)), \nonumber$
        \STATE $  \mathbf Z_2^{skip}, \mathbf Z_2 = \text{DS-Block}_{32\xrightarrow{}64}^{(2)}(\mathbf Z_2,  \mathbf{V}(t)), \nonumber$
        \STATE $    \mathbf Z_3^{skip}, \mathbf Z_3 = \text{DS-Block}_{64\xrightarrow{}128}^{(3)}(\mathbf Z_3,  \mathbf{V}(t)), \nonumber$
        \STATE $  \mathbf Z_3 = \text{US-Block}_{128\xrightarrow{}128}^{(3)}(\mathbf Z_3,\mathbf Z_3^{skip}, \mathbf{V}(t)), \nonumber$
        \STATE $    \mathbf Z_2 = \text{US-Block}_{256\xrightarrow{}64}^{(2)}(\mathbf Z_3, \mathbf Z_2^{skip}, \mathbf{V}(t)), \nonumber$
        \STATE $   \mathbf Z_1 = \text{US-Block}_{128\xrightarrow{}32}^{(1)}(\mathbf Z_2, \mathbf Z_1^{skip}, \mathbf{V}(t)), \nonumber$
        \STATE $   \mathbf Z_{out} =\text{D-Block}^{o}_{64\xrightarrow{}1}(\mathbf Z_1)/\beta(t). \nonumber$
     \STATE \textbf{Output: $ \mathbf Z_{out} \in\mathbb R^{B\times 1\times H\times W}$}.
	\end{algorithmic}
\end{Architecture}


The $\text{DS-Block}_{C_{in}\xrightarrow{}C_{out}}^{(i)}$($i=1,2,3$, $C_{out}=32,64,128$, respectively) are constructed according to Architecture \ref{al:score_dsblock}. The $\text{US-Block}_{C_{in}\xrightarrow{}C_{out}}^{(i)}$ ($i=1,2,3$, $C_{out}=32,64,128$, respectively) are constructed according to Architecture \ref{al:score_usblock}. 
Finally, the output decoding block $ \text{D-Block}^{o}_{64\xrightarrow{}1}$ generates  1-channel output.

\begin{Architecture}
	\caption{$\text{DS-Block}_{C_{in}\xrightarrow{}C_{out}}^{(i)}$}
	\label{al:score_dsblock}
	\begin{algorithmic}[1]  
		\STATE \textbf{Input:} $\mathbf Z^{block}_{in} \in\mathbb R^{B\times C_{in}\times H\times W}$, $\mathbf{V}(t)\in\mathbb R^{B\times 512}$.
        \STATE $ \mathbf Z_1 = \text{SiLU}(\text{GroupNorm}_{C_{out}/2} (\text{Conv2D}_{C_{out}}(\mathbf Z^{block}_{in}))), $
        \STATE $\mathbf Z_1 = \text{SiLU}(\text{GroupNorm}_{C_{out}/2} (\text{Conv2D}_{C_{out}}(\mathbf Z_1))),$
        \STATE $   \mathbf Z_1 = \mathbf Z_1 + \text{Dense}_{512\xrightarrow{}C_{out}}(\mathbf V(t)),$
        \STATE $   \mathbf Z_1 = \text{SiLU}(\text{GroupNorm}_{C_{out}/2} (\text{Conv2D}_{C_{out}}(\mathbf Z_1 ))),$
        \STATE $  \mathbf Z^{block,skip}_{out} = \text{SiLU}(\text{GroupNorm}_{C_{out}/2} (\text{Conv2D}_{C_{out}}(\mathbf Z_1 ))), \nonumber$\STATE $ \mathbf Z^{block}_{out} = \text{SiLU}((\text{Conv2D}_{C_{out}, /2}(\mathbf Z^{block}_{out}))). \nonumber$
         \STATE \textbf{Output:} $ \mathbf Z^{block,skip}_{out} \in\mathbb R^{B\times C_{out}\times H\times W}$, $ \mathbf Z^{block}_{out} \in\mathbb R^{B\times C_{out}\times H/2\times W/2}$.
	\end{algorithmic}
\end{Architecture}

\begin{Architecture}
	\caption{$\text{US-Block}_{C_{in}\xrightarrow{}C_{out}}^{(i)}$}
	\label{al:score_usblock}
	\begin{algorithmic}[1]  
		\STATE \textbf{Input:} $\mathbf Z^{block}_{in} \in\mathbb R^{B\times C_{in}\times H\times W}$, $\mathbf X^{block,skip}_{in} \in\mathbb R^{B\times C_{in}\times 2H\times 2W}$, $\mathbf{V}(t)\in\mathbb R^{B\times 512}$.
        \STATE $   \mathbf Z_1 = \text{SiLU}(\text{GroupNorm}_{C_{out}} (\text{Conv2D}_{2C_{out}}(\mathbf Z^{block}_{in}))), $
        \STATE $  \mathbf Z_1 = \text{SiLU}(\text{GroupNorm}_{C_{out}} (\text{Conv2D}_{2C_{out}}(\mathbf Z_1))), \nonumber$
        \STATE $   \mathbf Z_1 = \mathbf Z_1 + \text{Dense}_{512\xrightarrow{}2C_{out}}(\mathbf V(t)), \nonumber$
        \STATE $   \mathbf Z_1 = \text{SiLU}(\text{GroupNorm}_{C_{out}} (\text{Conv2D}_{2C_{out}}(\mathbf Z_1))), \nonumber $
        \STATE $    \mathbf Z_1 = \text{SiLU}(\text{GroupNorm}_{C_{out}} (\text{Conv2D}_{2C_{out}}(\mathbf Z_1))), \nonumber $
        \STATE $ \mathbf Z_1 = \text{SiLU}((\text{Conv2D}_{C_{out}, \times 2}(\mathbf Z_1)))), \nonumber$
        \STATE $\mathbf Z^{block}_{out} = \text{concat}[\mathbf Z_1,\mathbf Z^{block, skip}_{in} ]. \nonumber$
         \STATE \textbf{Output:} $ \mathbf Z^{block}_{out} \in\mathbb R^{B\times 2C_{out}\times 2H\times 2W}$.
	\end{algorithmic}
\end{Architecture}

\begin{Architecture}
	\caption{$ \text{D-Block}^{o}_{64\xrightarrow{}1}$}
	\label{al:score_usblock}
	\begin{algorithmic}[1]  
		\STATE \textbf{Input:} $\mathbf Z^{block}_{in} \in\mathbb R^{B\times 1\times H\times W}.$ 
        \STATE $        \mathbf Z_1 = \text{SiLU}((\text{Conv2D}_{64}(\mathbf Z^{block}_{in}))), \nonumber $
        \STATE $  \mathbf Z^{block}_{out} = \text{Conv2D}_{1}(\text{SiLU}((\text{Conv2D}_{64}(\mathbf Z_1)))). \nonumber$
         \STATE \textbf{Output:} $ \mathbf Z^{block}_{out} \in\mathbb R^{B\times 1\times H\times W}$.
	\end{algorithmic}
\end{Architecture}

\subsubsection{Neural network for baselines} \label{Autoencoder for baselines}
The architecture is used for DIS and BPS. Denote the input tensor as $\mathbf X\in\mathbb R^{B\times C\times H\times W}$ and the output tensor as $\mathbf X_{out}\in\mathbb R^{B\times U\times H\times W}$. The network is cascaded by encoding blocks $\text{E-Block}_{C_{in}\xrightarrow{}C_{out}}^{(i)}$ and decoding blocks $\text{D-Block}^{(i)}_{C_{in}\xrightarrow{}C_{out}}$, whose operations are the same as those in Architecture \ref{al:e-block-12} and \ref{al:d-block-12}, respectively. The whole architecture is in Architecture \ref{al:ae_baseline}. 
\begin{Architecture}
	\caption{Autoencoder for baselines}
	\label{al:ae_baseline}
	\begin{algorithmic}[1]  
		\STATE \textbf{Input:} $\mathbf X\in\mathbb R^{B\times C\times H\times W}.$ 
        \STATE $\mathbf X_1 = \text{E-Block}_{16\xrightarrow{}32}^{(2)}(\text{E-Block}_{C\xrightarrow{}16}^{(1)}(\mathbf X)),$
        \STATE $ \mathbf X_1 = \text{E-Block}_{64\xrightarrow{}128}^{(4)}(\text{E-Block}_{32\xrightarrow{}64}^{(3)}(\mathbf X_1)),$
        \STATE $\mathbf X_1 = \text{D-Block}_{128\xrightarrow{}64}^{(3)}(\text{D-Block}_{128\xrightarrow{}128}^{(4)}(\mathbf X_1)),$
        \STATE $   \mathbf X_1 = \text{D-Block}_{32\xrightarrow{}16}^{(1)}(\text{D-Block}_{64\xrightarrow{}32}^{(2)}(\mathbf X_1)), \nonumber$
        \STATE $   \mathbf X_{out} =\text{D-Block}^{(0)}_{16\xrightarrow{}U}(\mathbf X_1). \nonumber$
         \STATE \textbf{Output:} $\mathbf X_{out}\in\mathbb R^{B\times U\times H\times W}$.
	\end{algorithmic}
\end{Architecture}


\subsubsection{Score model for \ac{pdpnp}} \label{Score model for pdpnp}
The architecture is modified from Architecture \ref{al:score_ldpnp}, considering that the dimensions of input and output channels are different in the pixel and latent diffusion. Denote the input and output tensor as $\mathbf X, \mathbf X_{out} \in\mathbb R^{B\times C\times H\times W}$. The blocks $\text{DS-Block}_{C_{in}\xrightarrow{}C_{out}}^{(i)}$ and $\text{US-Block}_{C_{in}\xrightarrow{}C_{out}}^{(i)}$  ($i=1,2,3,4$, $C_{out}=32,64,128,256$, respectively) are cascaded as in Architecture \ref{al:score_pdpnp}. Operations inside $\text{DS-Block}_{C_{in}\xrightarrow{}C_{out}}^{(i)}$ and $\text{US-Block}_{C_{in}\xrightarrow{}C_{out}}^{(i)}$ can be found in Architecture \ref{al:score_dsblock} and \ref{al:score_usblock}, respectively. 

\begin{Architecture}
	\caption{Score model for \ac{pdpnp}}
	\label{al:score_pdpnp}
	\begin{algorithmic}[1]  
		\STATE \textbf{Input:} $\mathbf X \in\mathbb R^{B\times C\times H\times W}$, $t\in\mathbb R^B$.
        \STATE $ \mathbf V_p(t) = \text{concat}[\sin(2\pi t\mathbf W),\cos(2\pi t\mathbf W) ],$
        \STATE $    \mathbf V(t)=\text{SiLU}(\text{Dense}_{512\xrightarrow{}512} (\mathbf V_p(t))), \nonumber$
        \STATE $  \mathbf X_1^{skip}, \mathbf X_1 = \text{DS-Block}_{C\xrightarrow{}32}^{(1)}(\mathbf X,  \mathbf{V}(t)), \nonumber$
        \STATE $  \mathbf X_2^{skip}, \mathbf X_2 = \text{DS-Block}_{32\xrightarrow{}64}^{(2)}(\mathbf X_2,  \mathbf{V}(t)), \nonumber$
        \STATE $    \mathbf X_3^{skip}, \mathbf X_3 = \text{DS-Block}_{64\xrightarrow{}128}^{(3)}(\mathbf X_3,  \mathbf{V}(t)), \nonumber$
        \STATE $  \mathbf X_4^{skip}, \mathbf X_4 = \text{DS-Block}_{128\xrightarrow{}256}^{(3)}(\mathbf X_3,  \mathbf{V}(t)), \nonumber$
        \STATE $  \mathbf X_4 = \text{US-Block}_{256\xrightarrow{}256}^{(4)}(\mathbf X_4,\mathbf X_4^{skip}, \mathbf{V}(t)), \nonumber$
        \STATE $  \mathbf X_3 = \text{US-Block}_{512\xrightarrow{}128}^{(3)}(\mathbf X_4,\mathbf X_3^{skip}, \mathbf{V}(t)), \nonumber$
        \STATE $    \mathbf X_2 = \text{US-Block}_{256\xrightarrow{}64}^{(2)}(\mathbf X_3, \mathbf X_2^{skip}, \mathbf{V}(t)), \nonumber$
        \STATE $   \mathbf X_1 = \text{US-Block}_{128\xrightarrow{}32}^{(1)}(\mathbf X_2, \mathbf X_1^{skip}, \mathbf{V}(t)), \nonumber$
        \STATE $   \mathbf X_{out} =\text{D-Block}^{o}_{64\xrightarrow{}C}(\mathbf X_1)/\beta(t). \nonumber$
     \STATE \textbf{Output: $ \mathbf Z_{out} \in\mathbb R^{B\times C\times H\times W}$}.
	\end{algorithmic}
\end{Architecture}

\subsection{Implementation details}\label{appendix_B}
Our codes are written in PyTorch. All gradients are computed using the built-in automatic differentiation. In the following,  we use ${\boldsymbol{\chi}} \in \mathbb{C}^{N_x \times N_y}$ as a shorthand notation for the reshaped contrast vector $\text{diag}({\boldsymbol{\chi}}(\boldsymbol{\epsilon_r}, \boldsymbol{\sigma_e}))$.
\subsubsection{Occam inversion}
Occam inversion generates smooth boundary targets by penalizing the $\ell_2$ norm of spatial gradients. In our loss function, the unknown parameter is ${\boldsymbol{\chi}}$, and the data fidelity term is normalized by the $\ell_2$ norm of measurements. For the MNIST and Fashion-MNIST experiments, we set the smoothness regularization coefficient to 0.3 and performed 400 iterations of L-BFGS \cite{liu1989limited} (learning rate = 0.05). In the ATLAS example, we apply the multiplicative regularization \cite{zhang2019three} and run 400 iterations of ADAM \cite{kingma2014adam} (learning rate = 0.01). The regularization coefficient in each iteration is multiplied by measurement \ac{rmse} in each iteration,  with an initial coefficient being 20.
\subsubsection{TV-ADMM}
TV penalizes the $\ell_1$ norm of spatial gradients. In our loss function, the unknown parameter is ${\boldsymbol{\chi}} $, with the data fidelity term being normalized by the $\ell_2$ norm of measurements. In the MNIST and Fashion-MNIST examples, we set the TV coefficient to 5 and perform 20 ADMM iterations. Within each ADMM iteration, the quadratic loss function containing data fidelity is minimized by 20 steps of L-BFGS (learning rate = 0.1). In the ATLAS example, the TV coefficient is set to 1.5, and we also run 20 iterations of ADMM iterations. Within each ADMM iteration, the quadratic loss associated with data fidelity is minimized by 20 steps of ADAM (learning rate = 0.005).
\subsubsection{DIS}
The neural network architecture is presented in Appendix \ref{Autoencoder for baselines}. It takes modified measurements as input $\mathbf X\in\mathbb R^{B\times 2 N_f \times N_x\times N_y}$ and electrical property maps as output $\mathbf X_{out}\in\mathbb R^{B\times U\times N_x\times N_y}$, respectively. In training, the measurements are simulated from permittivity maps using (\ref{original_forward_model}), the former being inputs and the latter being labels. The training loss is the mean squared error between labels and predictions. The dataset for training is preprocessed as follows. We first reshape the measurements to $B\times 2 N_f \times N_T \times N_R $, where the channels are concatenated by the real and imaginary parts  at different frequency points. Then, bilinear interpolation is performed on the reshaped measurement tensor to produce the input tensor of size $B\times 2N_f \times N_x\times N_y$. The output is ${\boldsymbol{\chi}} $.  For the MNIST and Fashion-MNIST experiments,  $N_f=2$, $N_x=N_y=64$, and the output has one channel  $U=1$ representing the real part of ${\boldsymbol{\chi}} $. The model is trained for 20 epochs using a batch size of 256 and a exponentially decaying learning rate with the initial value of 0.0005 and a multiplicative factor of 0.99. In the ATLAS example, $N_f=2$, $N_x=N_y=96$, and the output has two channels $U=2$ that represent the real and imaginary part of  ${\boldsymbol{\chi}} $. The model is trained for 70 epochs using a batch size of 256 and a exponentially decaying learning rate with the initial value of 0.0005 and a multiplicative factor of 0.99. The ATLAS example takes more epochs because the inverse mapping from measurements to properties is more difficult to learn due to the complicated multiple scattering effects.
\subsubsection{BPS}
We apply the multiple-frequency back-projection \cite{guo2021physics} -- a classical linear imaging method that approximates the total field with the incident field -- to generate input data from the measurement. The neural network architecture is presented in Appendix \ref{Autoencoder for baselines}.
It takes back-projection results as input, $\mathbf X\in\mathbb R^{B\times C\times N_x\times N_y}$, and electrical property maps as output $\mathbf X_{out}\in\mathbb R^{B\times U\times N_x\times N_y}$, respectively.  The training loss is the mean squared error between the predicted and true electrical property maps. For the MNIST and Fashion-MNIST experiments,  we set $N_x=N_y=64$ and $C=U=1$ (for real-valued property maps). The model is trained for 100 epochs using a batch size of 256 and an exponentially decaying learning rate (initial value 0.0005, and multiplicative factor 0.99). For the ATLAS experiment, $N_x=N_y=96$ and $C=U=2$ (for complex-valued property maps). Since the dataset is larger than MNIST and Fashion-MNIST, we train for 70 epochs. Other settings are the same as those in the MNIST and Fashion-MNIST experiments. 
\subsubsection{GMR}\label{gmr}
It solves for the latent representation of the electrical properties via deterministic optimization. This process consists of two main steps: first, training an encoding model to learn the latent representation; and second, optimizing the latent variables to minimize the data fidelity term.

We use the neural network architecture detailed in Appendix \ref{Autoencoder for latent representation}  to reparameterize the property maps. The input and output has the same shape, $\mathbf X, \mathbf X_{out} \in\mathbb R^{B\times C\times N_x\times N_y}$. Here, the channels represent the real and imaginary components of the complex maps, with each channel normalized independently between $-1$ and $1$. The training loss is the ELBO that includes a small KL divergence term \cite{kingma2013auto,doersch2016tutorial}. All models are trained for 400 epochs using a batch size of 256 and an exponentially decaying learning rate with a decaying multiplicative factor 0.99. For the MNIST and Fashion-MNIST experiments, $N_x=N_y=64$ and $C=1$ (for real-valued property maps), with a KL divergence coefficient of 0.02, the initial learning rate of 0.0008. For the ATLAS experiment, $N_x=N_y=96$ and $C=2$ (for complex-valued property maps), with a KL divergence coefficient of 0.03, and the initial learning rate of 0.0005.

After the autoencoder is trained, we integrate the decoder into the optimization \cite{10332205} and solves for the latent variables using ADAM. The loss function consists of a data fidelity term normalized by the $\ell_2$ norm of measurements and a regularization term that penalizes the $\ell_2$ norm of the latent variables. The maximum step number of ADAM is set to 500. For the MNIST and Fashion-MNIST experiments, the learning rate is fixed at 0.08, and the initial latent variables are drawn from a standard Gaussian distribution. The regularization coefficient is 0.005. For the ATLAS experiment, we adopt a cosine annealing learning rate, with a starting value of 0.8,  minimum learning rate of 0.1, and a full cycle iteration number 1000.  The initial latent variables are set to the encoded latent variable of the average property map computed from the entire training dataset, and the regularization coefficient is 0.08.
\subsubsection{\ac{pdpnp}}\label{pdpnp_details}
The neural network's architecture is detailed in Appendix \ref{Score model for pdpnp}. The input and output has the same shape, $\mathbf X, \mathbf X_{out} \in\mathbb R^{B\times C\times N_x\times N_y}$. The channels represent the real and imaginary components of the complex maps, with each channel normalized independently between $-1$ and $1$. As suggested by \cite{songscore,song2022solving}, we set $f(t)=0$ and $g(t)=\sigma_d^t$ in (\ref{sde_fwd}). In this case, the variance of the transition distribution is 
\begin{equation}\label{beta_square}
    \beta^2(t) = \frac{1}{2\log\sigma_d}(\sigma_d^{2t}-1).
\end{equation}

We set $\sigma_d=20$. All models for different experiments are trained for 400 epochs using a batch size of 256 and an exponentially decaying learning rate (initial value 0.00008, and multiplicative factor 0.99). For the MNIST and Fashion-MNIST experiments, $C=1$, and $N_x=N_y= 64$.  For the ATLAS experiment, $C=2$, and $N_x=N_y= 96$. 

Recall that we define the following parameters: the number of the  two-step loops $N_k$, the number of iterations $N_\tau$ in the likelihood step, the number of iterations $N_t$ in the prior step, and the number of \ac{mmse} samples $M$. For all experiments, we set $N_k=20$, $N_\tau=120$, $N_t=500$, and $M=5$ (due to high computational cost). To balance the scales between the gradient and noise terms, we introduce a weighting coefficient $\alpha_n$ to the gradient term in (\ref{discrete_pcs}). The update equation becomes
\begin{equation}\label{discrete_pcs_modified}
\begin{aligned}
{\boldsymbol{\chi}}[n+1]&= \alpha_n \eta_k^2(1-r)\nabla_{{ {\boldsymbol{\chi}}}[n]}\mathcal L({\boldsymbol{\chi}}[n], \mathbf d_{obs}) + r {\boldsymbol{\chi}}[n] \\& + (1-r) \hat{\boldsymbol{\chi}}_k  + \eta_k \sqrt{(1-r^2)}\mathbf n,
\end{aligned}
\end{equation}
with 
\begin{equation}
\alpha_n = \alpha_0 \frac{\| {\boldsymbol{\chi}}[n] \|_2^2}{\eta_k^2(\|\nabla_{{ {\boldsymbol{\chi}}}[n]}\mathcal L({\boldsymbol{\chi}}[n], \mathbf d_{obs}) \|^2_2 + 0.001)},
\end{equation}
The initial image ${\boldsymbol{\chi}}[0]$ is drawn from a Gaussian distribution with a mean equal to the homogeneous background. We set $\gamma = 0.015\eta_k^2$ in $r=e^{-\gamma/\eta_k^2}$. For the prior step, 
 the initial time is computed according to (\ref{beta_square}) and $t_{N_t-1} = \beta^{-1}(\eta_k)$:
\begin{equation}
t_{N_t-1} =\frac{\log\left(1 + 2\,\eta_k\,\log\sigma_d\right)}{2\,\log\sigma_d} \bigg| _{\sigma_d=20}.
\end{equation}
In addition, we set $\varepsilon_t = 0.001$ when computing the time interval $\delta=\frac{t_{N_t}-\varepsilon_t}{N_t}$.

For the MNIST and Fashion-MNIST experiments, we set $\alpha_0 = 0.3$,  $\eta_k=0.4$ for $k=0,...,4$; and for $k=5,...,19$,  $\eta_k$ decays uniformly on a logarithmic scale from $0.4$ to $0.1$. In addition, since the transmitters and receivers are outside the imaging domain $D$, we do not apply masks in the likelihood step. For the ATLAS experiment, we set $\alpha_0 = 1$, and $\eta_k$ decays uniformly on a logarithmic scale from $0.2$ to $0.02$ for $k=0,...,19$. In each likelihood step (\ref{discrete_pcs_modified}), the gradients in regions near the sensors are set to zero by applying a circular mask with a radius of 0.12m.
\subsubsection{\ac{ldpnp}}
The neural network architecture of the autoencoder and score model is presented in Appendix \ref{Autoencoder for latent representation}  and \ref{Score model for ldpnp}, respectively. The setup for training the autoencoder is detailed in Appendix \ref{gmr}. The input and output of the score model in latent space  is $\mathbf X, \mathbf X_{out} \in\mathbb R^{B\times 1 \times N_u\times N_v}$, respectively. The settings for training the score model is the same as those in Appendix \ref{pdpnp_details}.  For the MNIST and Fashion-MNIST experiments,  $N_u=N_v= 16$.  For the ATLAS experiment,  $N_u=N_v= 24$. 

We set $N_k=20$,  $N_t=500$,  $N_\tau=120$, and $M=5$. The update equation for the likelihood step is
\begin{equation}\label{discrete_pcs_modified}
\begin{aligned}
{\mathbf z}[n+1]&= \alpha_n \eta_k^2(1-r)\nabla_{{ {\mathbf z}}[n]}\mathcal L(\mathbf {\mathbf z}[n], \mathbf d_{obs}) + r {\mathbf x}[n] \\& + (1-r) \hat{\mathbf z}_k  + \eta_k \sqrt{(1-r^2)}\mathbf n.
\end{aligned}
\end{equation}
with 
\begin{equation}
\alpha_n = \alpha_0 \frac{\| \mathbf z[n] \|_2^2}{\eta_k^2(\|\nabla_{{ {\mathbf z}}[n]}\mathcal L(\mathbf {\mathbf z}[n], \mathbf d_{obs}) \|^2_2 + 0.001)},
\end{equation}
where  $\mathbf z$  denotes the latent domain, and the initial  $\mathbf z[0]$ is drawn from the Gaussian distribution. For all experiments, we set $\alpha_0 = 1$.

For the MNIST and Fashion-MNIST experiments, $\eta_k=0.4$ for $k=0,...,4$; and for $k=5,...,19$,  $\eta_k$ decays uniformly on a logarithmic scale from $0.4$ to $0.1$. For the ATLAS experiment,  $\eta_k$ decays uniformly on a logarithmic scale from $1$ to $0.03$ for $k=0,...,19$. Other settings are kept the same with those in Appendix \ref{pdpnp_details}.

\bibliographystyle{ieeetr}
\bibliography{IEEEexample_v2}

\end{document}